\documentstyle[12pt]{article}

\font\blackboard=msbm10 at 12pt
\font\blackboards=msbm7
\font\blackboardss=msbm5
\newfam\black
\textfont\black=\blackboard
\scriptfont\black=\blackboards
\scriptscriptfont\black=\blackboardss

\newcommand{\junk}[1]{}

\newcommand{\ba}{\begin{array}}
\newcommand{\ea}{\end{array}}
\newcommand{\be}{\begin{equation}}
\newcommand{\ee}{\end{equation}}
\newcommand{\bea}{\begin{eqnarray}}
\newcommand{\eea}{\end{eqnarray}}
\newcommand{\beas}{\begin{eqnarray*}}
\newcommand{\eeas}{\end{eqnarray*}}

\def\identity{{\rlap{1} \hskip 1.6pt \hbox{1}}}

\def\laplace{{\kern1pt\vbox{\hrule height 1.2pt\hbox{\vrule width
1.2pt\hskip
  3pt\vbox{\vskip 6pt}\hskip 3pt\vrule width 0.6pt}\hrule height
  0.6pt}
  \kern1pt}}
\def\scriptlap{{\kern1pt\vbox{\hrule height 0.8pt\hbox{\vrule width
  0.8pt
  \hskip2pt\vbox{\vskip 4pt}\hskip 2pt\vrule width 0.4pt}\hrule height
  0.4pt}
  \kern1pt}}

\def\roughly#1{\raise.3ex\hbox{$#1$\kern-.75em\lower1ex\hbox{$\sim$}}}

\def\str{{\bf STr} \,}
\def\sym{{\rm Sym} \,}

\def\tr{{\bf Tr} \,}

\newcommand{\NP}{{\em Nucl.\ Phys.\ }}

\newcommand{\PL}{{\em Phys.\ Lett.\ }}
\newcommand{\PR}{{\em Phys.\ Rev.\ }}

\newcommand{\PRL}{{\em Phys.\ Rev.\ Lett.\ }}

\newcommand{\gone}[1]{}
\begin{document}
\pagestyle{plain}
\setcounter{page}{1}

\baselineskip16pt

\begin{titlepage}

\begin{flushright}
HUTP-00/A012\\
{\tt hep-th/0005145}
\end{flushright}
\vspace{8 mm}

\begin{center}

{\Large \bf Matrix Strings in Weakly Curved Background Fields\\}

\end{center}

\vspace{7 mm}

\begin{center}

{\bf Ricardo Schiappa}

\vspace{3mm}
{\small \sl Department of Physics} \\
{\small \sl Harvard University} \\
{\small \sl Cambridge, MA 02138, U.S.A.} \\

\vspace{3mm}

{\small \tt ricardo@lorentz.harvard.edu} \\

\end{center}

\vspace{8 mm}

\begin{abstract}
We investigate further the recent proposal for the form of the Matrix
theory action in weak background fields. We perform DVV reduction 
to the multiple $D0$--brane action in order to find the Matrix string theory 
action for multiple fundamental strings in curved but weak NS--NS and R--R 
backgrounds. This matrix sigma model gives a definite prescription on how 
to deal with R--R fields with an explicit spacetime dependence in Type II 
string theory. We do this both via the $9-11$ flip and the chain of $T$ 
and $S$ dualities, and further check on their equivalence explicitly. In 
order to do so, we also discuss the implementation of $S$--duality in 
the operators of the 2--dimensional world--volume supersymmetric gauge 
theory describing the Type IIB $D$--string. We compare the result to the 
known Green--Schwarz sigma model action (for one string), and use 
this comparison in order to discuss about possible, non--linear background 
curvature corrections to the Matrix string action (involving many strings), 
and therefore to the Matrix theory action. We illustrate the nonabelian 
character of our action with an example involving multiple fundamental 
strings in a non--trivial R--R flux, where the strings are polarized into a 
noncommutative configuration. This corresponds to a dielectric type of 
effect on fundamental strings.
\end{abstract}

\vspace{1cm}
\begin{flushleft}
April 2000
\end{flushleft}
\end{titlepage}
\newpage

\section{Introduction}

The five known superstring theories as well as the low--energy 11--dimensional 
supergravity are known to be related through a web of dualities, and it is 
believed that all these theories are simply different limits of an underlying 
11--dimensional quantum theory known as $M$--theory, whose fundamental degrees 
of freedom are as yet unknown, but that can be defined as the strong coupling 
limit of Type IIA string theory \cite{Witten-various, Polchinski}. Let us first 
recall that $M$--theory compactified on a circle is described by Type IIA 
string theory at finite string coupling. It is by now a well known conjecture 
that $M$--theory compactified on a lightlike circle admits a nonperturbative 
description in terms of the degrees of freedom of a collection of $D0$--branes 
\cite{Witten-bound, BFSS, bss, Susskind-DLCQ, Sen-DLCQ, Seiberg-DLCQ}. 

Matrix theory encodes a great deal of information about the structure of 
both $M$--theory and 11--dimensional supergravity (some reviews are 
\cite{WT-Trieste, Douglas-lectures, Wati-Matrix}). One knows how to identify 
supergravitons, membranes and fivebranes in Matrix theory \cite{BFSS, 
bss, grt}, and the interactions between these objects in Matrix theory 
have been found to agree with supergravity in a variety of situations. 
In particular, for general Matrix configurations, it was shown in 
\cite{Dan-Wati-2, Mark-Wati-3} that the supergravity potential between 
an arbitrary pair of $M$--theory objects arising from the exchange of 
quanta with zero longitudinal momentum is exactly reproduced by terms 
in the one--loop Matrix theory potential. These results were also used to 
describe a formulation of Matrix theory in a general metric and 3--form 
background, via a matrix sigma model type of action \cite{Mark-Wati-3}. 
Such matrix sigma model actions had also been advocated for earlier in 
\cite{Douglas-talk}. A different type of approach to the 3--form 
background is, {\it e.g.} \cite{Chu-Ho-Li}.

A question that naturally arises is that if we have a formulation of 
Matrix theory in curved background fields, that should somehow yield a 
matrix formulation of Type II string theory in curved background fields, 
and in particular in the presence of R--R fields. Moreover, due to the second 
quantized nature of the Matrix theory formalism, we should be able to 
obtain in this way a description of multiple interacting strings in both 
NS--NS and R--R curved backgrounds. This would be quite interesting, as 
even for a single fundamental superstring the action in a general background 
including arbitrary R--R fields is not yet well understood.

Due to the relation between Type IIA string theory and $M$--theory, it is 
possible to construct a matrix theory formulation of superstring 
theory which is known as matrix string theory \cite{Motl, 
Banks-Seiberg, Dijkgraaf-Verlinde-Verlinde, Banks-Motl, 
Dijkgraaf-Verlinde-Verlinde-2}. Such a formulation is achieved once 
one understands toroidal compactifications of Matrix theory 
\cite{WT-compact}, for then the particular case of the ${\bf S}^{1}$ 
compactification will lead to the matrix formulation of the Type IIA 
superstring -- as $M$--theory compactified on a circle yields the IIA 
theory, where the IIA string is obtained from the wrapped 
$M2$--brane \cite{DHIS}. Recall that this matrix string theory is a 
supersymmetric gauge theory that not only contains all of the DLCQ IIA 
superstring theory, but also contains extra degrees of freedom which represent 
nonperturbative objects in string theory. These nonperturbative degrees 
of freedom represent the inclusion of $D$--brane states, and also give 
us a prescription to include nonperturbative corrections in calculations 
of diverse processes in perturbative string theory. 

Because we know a great deal about Type IIA string theory, matrix 
string theory is a very good laboratory to test Matrix theory. Of 
course ideally we would like to have a microscopic definition of 
$M$--theory which would be covariant and defined in arbitrarily curved 
backgrounds. But due to the nonabelian character of the theory 
such is not an easy goal. Information from the abelian limit of the 
theory may then prove to be of great value in trying to deal with such 
issues, and a precious source of information on this abelian limit 
is undoubtably the Type II theory. In flat space the matrix string 
theory action has been lifted from the cylinder to its branched 
coverings and a precise connection with the Green--Schwarz action in 
light--cone gauge has been achieved \cite{Bonelli-Bonora-Nesti-1, 
Bonelli-Bonora-Nesti-2, Bonelli-Bonora-Nesti-Tomasiello-1, 
Bonelli-Bonora-Nesti-Tomasiello-2, Bonelli, Paniak}, with the interesting 
result that the full moduli space of the IIA theory is recovered 
within matrix string theory only in the large $N$ limit. Scattering 
amplitudes have been reproduced within the matrix string formalism in 
\cite{Wynter-2, Giddings-Hacquebord-Verlinde}, for reviews on 
several issues see \cite{Dijkgraaf-Verlinde-Verlinde-2, Hacquebord}. 
More recently, the issue of a spacetime covariant formulation of matrix 
string theory has been addressed in \cite{Baulieu-Laroche-Nekrasov}, 
but this is a matter which is far from settled.

This paper concerns the generalization of matrix string theory when in the 
presence of weakly curved background fields. In particular, we want to 
address the question of how to describe multiple interacting strings 
in NS--NS and R--R curved backgrounds. Indeed, because it is known how to 
describe the linear couplings of Matrix theory to a curved 
11--dimensional background, we shall also be able to find the linear 
couplings of matrix string theory to a curved 10--dimensional background. 
This could be of great interest not only in trying to improve our 
knowledge of string theory in R--R backgrounds, but also when 
comparing to the IIA theory in the abelian limit we could expect for 
new information on how to construct Matrix theory in a general curved 
background. In summary, we are looking for a matrix string sigma model 
type of action,

\bea
{\cal S} &=& {1 \over 2\pi} \int d\sigma d\tau\; ( {1\over 2} 
g_{\mu\nu}^{IIA} (X) I^{\mu\nu}_{g} + \phi (X) I_{\phi} + B_{\mu\nu} (X) 
I^{\mu\nu}_{s} + \widetilde{B}_{\mu\nu\lambda\rho\sigma\tau} (X) 
I^{\mu\nu\lambda\rho\sigma\tau}_{5} 
\nonumber \\
&
& + 
C_{\mu} (X) I^{\mu}_{0} + \widetilde{C}_{\mu\nu\lambda\rho\sigma\tau\xi} (X) 
I^{\mu\nu\lambda\rho\sigma\tau\xi}_{6} + C_{\mu\nu\lambda} (X) 
I^{\mu\nu\lambda}_{2} + \widetilde{C}_{\mu\nu\lambda\rho\sigma} (X) 
I^{\mu\nu\lambda\rho\sigma}_{4} ),
\eea

\noindent
and we shall precisely explain in this paper how to construct this 
action by specifying both the $I$ tensor couplings as well as the 
inclusion of spacetime dependence in the (weak) background fields. The 
explicit form of all these tensor couplings is presented in section 4.2.

We shall begin in section 2 with a brief review of the work done for 
the case of Matrix theory in weak background fields \cite{Dan-Wati-2, 
Mark-Wati-3, Mark-Wati-4, Mark-Wati-5}. We shall recall that there is a 
definite proposal on how to supplement the flat space Matrix action 
with linear couplings between the background fields -- the supergraviton, 
the membrane and the fivebrane -- and the respective Matrix descriptions 
for the supergravity stress--energy tensor, membrane current and fivebrane 
current. Moreover we shall also recall that through the Sen--Seiberg 
limiting procedure, this action can be reduced to an action for multiple 
$D0$--branes in weakly curved Type IIA background fields. By $T$--duality 
this can be extended to any Type II $D$--brane. Then, in section 3 we 
present a brief review of the Dijkgraaf--Verlinde--Verlinde (DVV) reduction 
of Matrix theory to matrix string theory \cite{Dijkgraaf-Verlinde-Verlinde}, 
via both the so--called $9-11$ flip and also the $T$--$S$--$T$ chain 
of dualities. This will be of fundamental use in the sections that 
follows, as we shall be generalizing that procedure to the curved 
background situation.

In the following sections we perform the DVV reduction to the multiple 
$D0$--brane action in order to find the matrix string theory action for 
multiple fundamental strings in curved but weak NS--NS and 
R--R backgrounds. As we just said, this is a generalization of the work 
by DVV. These sections deals with a great deal of algebra, and we will 
be schematic in presenting our results. The matrix sigma model 
obtained in this way gives a definite prescription on how to deal with 
R--R fields with an explicit spacetime dependence in Type II string 
theory. Due to the nonabelian nature of the action, it also gives a 
second quantized description of Type II string theory in such 
backgrounds. We shall obtain the matrix string sigma model both via 
the $9-11$ flip (described in section 4) and the chain of $T$ and $S$ 
dualities (described in section 5), and further check their equivalence 
explicitly by obtaining the same results in both cases. In order to do 
so, we will need to discuss in section 5 the implementation of $S$--duality 
in the composite operators of the 2--dimensional world--volume supersymmetric 
gauge theory describing the Type IIB $D$--string. We shall obtain 
the $S$--duality transformations for the world--volume fields from the 
equivalence with the $9-11$ flip, and we shall see that these 
transformation properties are indeed quite simple, as should be expected.

In section 6 we compare the result to the known Green--Schwarz 
sigma model action (for one string) \cite{Fradkin-Tseytlin}. This is 
done by extracting the free string limit (the IR limit of the gauge 
theory) of the matrix string theory action. This will be a qualitative 
match only, as we shall not construct the precise lifting of the matrix 
string action to the Green--Schwarz action. We then use this comparison in 
order to discuss about possible, non--linear background curvature corrections 
to the matrix string action (involving many strings), and therefore to the 
Matrix theory action. Again this is a qualitative analysis, but it gives us 
further insight into the goal of constructing Matrix theory in arbitrary curved 
backgrounds. Then, in section 7, we briefly discuss the exponentiation 
of the noncommutative vertex operators we obtained in order to build 
coherent states of fundamental strings and so obtain the full 
non--linear matrix string sigma model. As such a construction is not 
clear at this stage, we turn to an illustration of the nonabelian character of 
our action with an example, namely multiple fundamental strings in a 
non--trivial R--R flux, where the strings are polarized into 
nonabelian configurations due to the background field. This means that 
Myers' dielectric effect for $D$--branes has an analogue for 
fundamental strings. We also speculate on a possible relation between 
this effect and string theory noncommutative background geometries, 
where this could provide a very interesting example of target space 
noncommutativity in the presence of R--R fields (as opposed to recent 
discussions of world--volume noncommutativity in the presence of NS--NS fields, 
{\it e.g.}, \cite{Douglas-lectures, Connes-Douglas-Schwarz, Lorenzo-Ricardo, 
Seiberg-Witten, Lorenzo}). We conclude in section 8 with some open problems 
for future research.

\section{Matrix Theory in Weakly Curved Backgrounds}

We begin with a short review of the results obtained for Matrix 
theory in weakly curved background fields \cite{Dan-Wati-2,Mark-Wati-3}, 
and also for the action of multiple $D0$--branes in weak Type IIA backgrounds 
\cite{Mark-Wati-4} as well as for the action of multiple $Dp$--branes in 
Type II weak background fields \cite{Mark-Wati-5}.

\subsection{Results for Matrix Theory}

In this section we briefly review the results in \cite{Dan-Wati-2,Mark-Wati-3} 
dealing with the construction of a Matrix theory action in weak 
$M$--theory backgrounds. As we shall see, due to the $9-11$ flip in the 
DVV construction of matrix string theory, we will have particular 
interest in the tensors that appear in this Matrix theory action. 

The proposal in question actually concerns the terms in the action of 
Matrix theory which are linear in the background fields 
\cite{Mark-Wati-3}. If we consider a general Matrix theory background, 
with metric $g_{IJ} = \eta_{IJ} + h_{IJ}$ and 3--form field $A_{IJK}$, 
then the linear effects of this background can be described by 
supplementing the flat space Matrix theory action,

\be
S_{Flat} = {1\over R} \int dt \; \tr \left( {1\over 2} D_{t} X^{i} 
D_{t} X^{i} - {1\over 2} \sum_{i<j} [X^{i}, X^{j}]^{2} - {1\over 
2} \Theta D_{t} \Theta + {1\over 2} \Theta \gamma^{i} [X_{i}, 
\Theta] \right),
\ee

\noindent
with additional linear coupling terms of the form,

\bea \label{mta}
S_{Weak} &=& \int dt \; \sum_{n=0}^{\infty} \; \sum_{i_{1}, \ldots, 
i_{n}} \; {1\over n!} \; \{ {1\over2} T^{IJ (i_{1} \cdots i_{n})} 
\partial_{i_{1}} \cdots \partial_{i_{n}} h_{IJ}(0) + J^{IJK (i_{1} 
\cdots i_{n})} \partial_{i_{1}} \cdots \partial_{i_{n}} A_{IJK}(0) 
\nonumber \\
&
& + 
M^{IJKLMN (i_{1} \cdots i_{n})} \partial_{i_{1}} \cdots 
\partial_{i_{n}} \widetilde{A}_{IJKLMN}(0) + {\rm Fermionic} \;\; 
{\rm Terms} \},
\eea

\noindent
where $\widetilde{A}$ is the dual 6--form field which satisfies at linear 
order,

\be
d \widetilde{A} = \star \; d A.
\ee

\noindent
The previous matrix expressions $T^{IJ(i_{1} \cdots i_{n})}$, 
$J^{IJK(i_{1} \cdots i_{n})}$ and $M^{IJKLMN(i_{1} \cdots i_{n})}$ are 
the Matrix theory forms of the multipole moments of the stress--energy 
tensor, membrane current and 5--brane current of 11--dimensional 
supergravity. Explicit forms for the bosonic parts of these moments 
were first given in \cite{Dan-Wati-2}, and those results were 
later extended to quadratic fermionic terms (and also some quartic fermionic 
terms) in \cite{Mark-Wati-3}. The complete results in \cite{Dan-Wati-2, 
Mark-Wati-3} are reproduced in the Appendix.

With these definitions the previous expressions yield a formulation 
of Matrix theory in a weak background metric to first order in 
$h_{IJ}$, the 3--form $A_{IJK}$, and all their higher derivatives. It 
was moreover argued in \cite{Mark-Wati-3} that if the Matrix theory 
conjecture is true in flat space, then this formulation must be 
correct at least to order ${\cal O}(\partial^{4}h,\partial^{4}A)$. 
It was also conjectured in that paper that this form may work to all 
orders in derivatives of the background fields, and in a general background. 
One should observe however that it is not known how to incorporate dependence 
of the background on the compact coordinate $X^{-}$.

\subsection{Results for Multiple D--branes}

We proceed by reviewing how the previous results can be used to 
construct actions for multiple $D0$--branes \cite{Mark-Wati-4} and in 
general for multiple $Dp$--branes \cite{Mark-Wati-5} in Type II string 
theory, in the approximation of weak background fields. Of particular 
interest to our goal in this paper is the case of the $D0$--brane 
action, due to the duality sequence in the DVV construction of matrix 
string theory and its associated $9-11$ flip.

To start, we shall recall from \cite{Mark-Wati-4} how one obtains the 
action for multiple $D0$--branes in background fields, as this will 
later prove its interest when we try to do the same for the matrix 
string action. We begin with $M$--theory on a background metric, 
$g_{IJ} = \eta_{IJ} + h_{IJ}$, in a frame where there is a compact 
coordinate $X^{-}$ of size $R$, which becomes lightlike in the flat 
space limit, $g_{IJ} \to \eta_{IJ}$. From the Sen--Seiberg limit 
\cite{Sen-DLCQ, Seiberg-DLCQ} we know that this theory can be 
described as a limit of a family of spacelike compactified theories. 
If we define an $\tilde{M}$--theory with background metric 
$\tilde{g}_{IJ} = \eta_{IJ} + \tilde{h}_{IJ}$, in a frame with a 
spacelike compact coordinate $X^{11}$ of size $R_{11}$, then the DLCQ 
limit of the original $M$--theory is found by boosting the 
$\tilde{M}$--theory along $X^{11}$, and then taking the limit $R_{11} \to 0$. 
Knowing the boost we can trivially Lorentz relate the metric $\tilde{g}_{IJ}$ 
in the $\tilde{M}$--theory with the metric $g_{IJ}$ in the $M$--theory. 
Moreover, in the DLCQ description the $M$--theory is in light--cone 
coordinates, $X^{\pm} = {1\over\sqrt{2}} (X^{0} \pm X^{11})$, and so 
it is easy to relate the metric $\tilde{g}_{IJ}$ to the light--cone 
metric $g_{IJ}$.

Of course our final goal is more than we have just obtained. We would 
like to relate the Type IIA string theory background fields to the 
DLCQ $M$--theory ones. But this is now straightforward. 
$\tilde{M}$--theory on a small spacelike circle of radius $R_{11}$ is 
known to be equivalent to Type IIA string theory with background fields 
given to leading order by,

\bea
h^{IIA}_{\mu\nu} &=& \tilde{h}_{\mu\nu} + {1\over2} \eta_{\mu\nu} 
\tilde{h}_{11 \; 11}, \nonumber \\
C_{\mu} &=& \tilde{h}_{11 \; \mu}, \nonumber \\
\phi &=& {3\over 4} \tilde{h}_{11 \; 11}.
\eea

\noindent
All we have left to do is to relate the $\tilde{h}_{IJ}$ metric to 
the $h_{IJ}$ one through the previously explained procedure. In order 
to describe nontrivial background antisymmetric tensor fields, one 
should also include the connections between the IIA background fields 
and the $M$--theory background 3--form field. The action for multiple 
$D0$--branes can now be obtained by direct comparison with the one for 
Matrix theory just described in the previous subsection. Indeed 
\cite{Mark-Wati-4}, one can first write the $D0$--brane action in terms of 
some unknown quantities coupling to the background fields. These quantities
will be denoted by $I_{x}$ and will couple linearly to each of the 
background fields, so that to leading order the action for $N$ 
$D0$--branes is written as:

\bea
{S}_{D0-branes} = {S}_{Flat} &+& \int dt
\sum_{n=0}^{\infty} {1\over n!} \{ {1\over2} (\partial_{k_{1}} \cdots 
\partial_{k_{n}} h^{IIA}_{\mu\nu}) I_{h}^{\mu\nu(k_{1} \cdots 
k_{n})} + (\partial_{k_{1}} \cdots \partial_{k_{n}} \phi) 
I_{\phi}^{(k_{1} \cdots k_{n})} \nonumber \\
&+& (\partial_{k_{1}} \cdots \partial_{k_{n}} C_{\mu}) 
I_{0}^{\mu(k_{1} \cdots k_{n})} + (\partial_{k_{1}} \cdots 
\partial_{k_{n}} \widetilde{C}_{\mu\nu\lambda\rho\sigma\tau\xi}) 
I_{6}^{\mu\nu\lambda\rho\sigma\tau\xi(k_{1} \cdots k_{n})} \nonumber 
\\
&+& (\partial_{k_{1}} \cdots \partial_{k_{n}} B_{\mu\nu}) 
I_{s}^{\mu\nu(k_{1} \cdots k_{n})} + (\partial_{k_{1}} \cdots 
\partial_{k_{n}} \widetilde{B}_{\mu\nu\lambda\rho\sigma\tau}) 
I_{5}^{\mu\nu\lambda\rho\sigma\tau(k_{1} \cdots k_{n})} \nonumber \\
&+& (\partial_{k_{1}} \cdots \partial_{k_{n}} C_{\mu\nu\lambda}) 
I_{2}^{\mu\nu\lambda(k_{1} \cdots k_{n})} + (\partial_{k_{1}} \cdots 
\partial_{k_{n}} \widetilde{C}_{\mu\nu\lambda\rho\sigma}) 
I_{4}^{\mu\nu\lambda\rho\sigma(k_{1} \cdots k_{n})} \}.
\eea

Replacing in this action the background fields of the Type IIA string 
theory by the background fields of DLCQ $M$--theory according to the 
previous relations, one can then compare the previous action for 
$D0$--branes to the Matrix theory action and deduce the expressions 
for the string theory couplings $I_{x}$. These are \cite{Mark-Wati-4}:

\bea
&&I^{00}_{h} = T^{++} + T^{+-} + (I^{00}_{h})_{8} + {\cal O}(X^{12}),
\nonumber \\
&&I^{0i}_{h} = T^{+i} + T^{-i} + {\cal O}(X^{10}),
\nonumber \\
&&I^{ij}_{h} = T^{ij} + (I^{ij}_{h})_{8} + {\cal O}(X^{12}), 
\nonumber \\
&&I_{\phi} = T^{++} - {1\over 3} ( T^{+-} + T^{ii} ) + 
(I_{\phi})_{8} + {\cal O}(X^{12}),
\nonumber \\
&&I^{0i}_{s} = 3 J^{+-i} + {\cal O}(X^{8}),
\nonumber \\
&&I^{ij}_{s} = 3 J^{+ij} - 3 J^{-ij} + {\cal O}(X^{10}),
\nonumber \\
&&I^{0}_{0} = T^{++},
\nonumber \\
&&I^{i}_{0} = T^{+i},
\nonumber \\
&&I^{0ij}_{2} = J^{+ij} + {\cal O}(X^{10}), 
\nonumber \\
&&I^{ijk}_{2} = J^{ijk} + {\cal O} (X^{8}),
\nonumber \\
&&I^{0ijkl}_{4} = 6 M^{+-ijkl} + {\cal O}(X^{8}), 
\nonumber \\
&&I^{ijklm}_{4} = - 6 M^{-ijklm} + {\cal O}(X^{10}), 
\nonumber \\
&&I^{0ijklmn}_{6} = S^{+ijklmn} + {\cal O}(X^{10}),
\nonumber \\
&&I^{ijklmnp}_{6} = S^{ijklmnp} + {\cal O}(X^{12}).
\eea

\noindent
By $T$--duality of background supergravity fields and $T$--duality of 
world--volume fields, the previous action for $N$ $D0$--branes can be 
transformed into an action for $N$ Type II $Dp$--branes, as was 
discussed in \cite{Mark-Wati-5}. This also allows for a discussion of 
nonabelian terms in the Born--Infeld action. For further discussion we 
refer the reader to the original references \cite{Mark-Wati-4, 
Mark-Wati-5}.

\section{Matrix String Theory}

According to the DVV formulation of matrix string theory 
\cite{Dijkgraaf-Verlinde-Verlinde}, one can reduce the Matrix theory 
action to an action for IIA matrix strings in two different ways. One way is 
by performing the so--called $9-11$ flip, where one exchanges the 
role of the $9^{th}$ and $11^{th}$ directions of $M$--theory. Another 
way is via a set of dualities on the background fields. Moreover, the 
coordinate flip should clearly be equivalent to this specific chain of 
dualities. In here, one starts by $T$--dualizing and then takes an $S$--duality 
followed by another $T$--duality. The starting point is the Type IIA theory, 
with $N_{11}$ yielding the $D$--particle number. After the $T$--duality along 
$R_{9}^{IIA}$ one reaches Type IIB, where $N_{11}$ now equals the 
$D$--string number. The Type IIB $S$--duality leads to $N_{11}$ equaling 
the $F$--string number, and the final $T$--duality along $R_{9}^{IIB}$ leads 
back to Type IIA, with $N_{11}$ now being equal to the $F$--string momenta.

In order to see that this exactly matches the simple $9-11$ flip on 
the compact coordinates, let us follow these dualities with a slightly 
greater detail \cite{Dijkgraaf-Verlinde-Verlinde}. If we compactify 
$M$--theory on ${\bf S}^{1}_{R_{9}} \times {\bf S}^{1}_{R_{11}}$, with 
$R_{11}$ the spacelike compact direction which becomes lightlike in 
the Sen--Seiberg limit, we will have the parameters,

\be
R_{11}=g_{s}\ell_{s}, \ \ \ \ \ \ \ \ell_{P}^{3}=g_{s}\ell_{s}^{3},
\ee

\noindent
and also $R_{9}$ for the remaining spacelike compact direction. The 
$9-11$ flip simply leads to the IIA theory with parameters,

\be
R_{9}=g'_{s}\ell_{s}, \ \ \ \ \ \ \ \ell_{P}^{3}=g'_{s}\ell_{s}^{3},
\ee

\noindent
where now the remaining spacelike compact direction is $R_{11}$. On 
the other hand, given our starting point and $T$--dualizing along 
$R_{9}^{IIA}$, one obtains the following Type IIB parameters,

\bea
g_{s}^{IIB} &=& {\ell_{s} \over R_{9}^{IIA}}\; g_{s}^{IIA} = {\ell_{s} \over 
R_{9}^{IIA}}\; {R_{11} \over \ell_{s}} = {R_{11} \over R_{9}^{IIA}}, \\
R_{9}^{IIB} &=& {\alpha' \over R_{9}^{IIA}} = {\ell_{s}^{2} \over 
R_{9}^{IIA}}.
\eea

\noindent
A further IIB $S$--duality leads to

\bea
{g'}_{s}^{IIB} &=& {1 \over g_{s}^{IIB}} = {R_{9}^{IIA} \over R_{11}}, \\
{R'}_{9}^{IIB} &=& {1 \over g_{s}^{IIB}} \; R_{9}^{IIB} = \left( 
{R_{11} \over R_{9}^{IIA}} \right)^{-1}\; {\ell_{s}^{2} \over 
R_{9}^{IIA}}.
\eea

\noindent
In the expressions above for the radius, recall that under 
$T$--duality it is the Einstein frame metric that is invariant. The 
string frame metric gets transformed with a $g_{s}$ factor. Finally, 
we finish the chain of dualities by $T$--dualizing back to the IIA 
theory along ${R'}_{9}^{IIB}$. We end up with the parameters,

\bea
{g'}_{s}^{IIA} &=& {\ell_{s} \over {R'}_{9}^{IIB}}\; {g'}_{s}^{IIB} = 
{R_{9} \over \ell_{s}}, \\
{R'}_{9}^{IIA} &=& {\ell_{s}^{2} \over {R'}_{9}^{IIB}} = R_{11},
\eea

\noindent
which are exactly the same as the ones obtained via the $9-11$ flip.

Given that, as we have just seen, the chain of dualities is equivalent 
to the $9-11$ flip, we shall now obtain the matrix string action from 
the Matrix action by following the most straightforward path, {\it i.e.}, 
we shall simply perform the flip to the Matrix theory action 
\cite{Dijkgraaf-Verlinde-Verlinde}. With the dimensionfull parameters 
made explicit, in order to produce the correct dimensions for the fields, 
the Matrix action in a flat  background is written as,

\be
S = \int dt\; {\bf Tr} \left( {1 \over 2R} \dot{X_{i}}\dot{X_{i}} + 
{R M_{P}^{6} \over 8\pi^{2}} \sum_{i < j} [X^{i},X^{j}]^{2} + 
{i M_{P}^{3} \over 4\pi} \theta^{T}\dot{\theta} - 
{R M_{P}^{6} \over 8\pi^{2}} \theta^{T} \gamma_{i} [X^{i},\theta] 
\right),
\ee

\noindent
with $R=2\pi\ell_{P}^{3}$ and $M_{P}$ is the Planck mass.

We further consider the theory compactified along the $9^{th}$ 
direction. Therefore, defining $\hat{R}_{9}={\alpha' \over R_{9}}$, one 
$T$--dualizes according to the standard procedure and obtains:

\bea
S' &=& \int dt\; {1 \over 2\pi\hat{R}_{9}} \int_{0}^{2\pi\hat{R}_{9}} 
d\hat{x}\; {\bf Tr}\; ( {1 \over 2R} \dot{X_{i}}\dot{X_{i}} + 
{1 \over 2R}(2\pi\alpha')^{2}\dot{A}^{2} + 
{R M_{P}^{6} \over 8\pi^{2}} \sum_{i<j} [X^{i},X^{j}]^{2} \nonumber \\
&
& - 
{R M_{P}^{6} \over 8\pi^{2}}(2\pi\alpha')^{2}(D_{\hat{x}}X^{i})^{2} + 
{i M_{P}^{3} \over 4\pi} \theta^{T}\dot{\theta} - 
{R M_{P}^{6} \over 8\pi^{2}} \theta^{T} \gamma_{i} [X^{i},\theta] -
i{R M_{P}^{6} \over 8\pi^{2}}(2\pi\alpha') \theta^{T} \gamma_{9} 
D_{\hat{x}}\theta ) . \nonumber
\eea

\noindent
The implementation of the $9-11$ flip is quite simple, as one 
just has to notice the change in parameters so that 
$R_{9}=g_{s}\ell_{s}$ and $\hat{R}_{9}={\ell_{s} \over g_{s}}$. 
Consequently,

\bea
S' &=& \int dt\; {g_{s} \over 2\pi\ell_{s}} \int_{0}^{2\pi{\ell_{s} \over 
g_{s}}} d\hat{x}\; {\bf Tr}\; ( {1 \over 2R} \dot{X_{i}}\dot{X_{i}} + 
{2\pi^{2}\ell_{s}^{4} \over R}\dot{A}^{2} + 
{R M_{P}^{6} \over 8\pi^{2}} \sum_{i<j} [X^{i},X^{j}]^{2} \nonumber \\
&
& - 
{1 \over 2} R M_{P}^{6} \ell_{s}^{4} (D_{\hat{x}}X^{i})^{2} + 
{i M_{P}^{3} \over 4\pi} \theta^{T}\dot{\theta} - 
{R M_{P}^{6} \over 8\pi^{2}} \theta^{T} \gamma_{i} [X^{i},\theta] -
{i R M_{P}^{6} \ell_{s}^{2} \over 4\pi} \theta^{T} \gamma_{9} 
D_{\hat{x}}\theta ) . \nonumber
\eea

One can rescale the world--sheet coordinates, from $(\hat{x},t)$ 
to $(\sigma,\tau)$, such that $0<\sigma<2\pi$ and so that the coordinates on 
the cylinder become dimensionless. For that one changes 
$\hat{x}={\ell_{s} \over g_{s}}\sigma$ (and therefore 
$D_{\hat{x}}={g_{s} \over \ell_{s}} D$ \footnote{Throughout, 
derivatives with no explicit subscript shall refer to the cylinder 
world--sheet index $\sigma$, {\it i.e.}, $D\equiv D_{\sigma}$ and 
$\partial \equiv \partial_{\sigma}$.}). We also have to rescale time 
on the world-sheet $t={\ell_{s}^{2} \over R}\tau$. Moreover, we shall 
deal with dimensionless background target fields such that they will 
be measured in string units, {\it i.e.}, rescale $(X,\theta)$ to 
$(\ell_{s}X,\ell_{s}\theta)$. All this done, we are left with the 
rescaled action,

\bea
S' &=& {1\over 2\pi} \int d\tau d\sigma\; {\bf Tr}\;
({1\over2}\dot{X_{i}}\dot{X_{i}} + 2\pi^{2}g_{s}^{2}\dot{A}^{2} + 
{1\over8\pi^{2}g_{s}^{2}} \sum_{i<j} [X^{i},X^{j}]^{2} \nonumber \\
&
& - 
{1\over2}(D X^{i})^{2} + {i\over 4\pi R_{9}} \theta^{T}\dot{\theta} - 
{1\over 8\pi^{2}g_{s} R_{9}} \theta^{T} \gamma_{i} [X^{i},\theta] -
{i\over 4\pi R_{9}} \theta^{T} \gamma_{9} D \theta ) . \nonumber
\eea

\noindent
In order to cast this action into a more familiar looking one, we simply 
have to perform one further rescaling of the background fermions, $\theta 
\to \sqrt{4\pi R_{9}} \theta$, and change the notation for the string 
coupling constant as $g_{s} \to {g_{s}\over 2\pi}$. Then the $9-11$ 
flip is concluded and we have obtained the DVV reduction of the Matrix 
theory action,

\bea
S &=& {1\over 2\pi} \int d\tau d\sigma\; {\bf Tr}\; ({1\over2} 
( (\dot{X_{i}})^{2} -(DX^{i})^{2} ) + {1\over 2g_{s}^{2}} \sum_{i<j} 
[X^{i},X^{j}]^{2} + {1\over2}g_{s}^{2}\dot{A}^{2} \nonumber \\
&
& + 
i( \theta^{T}\dot{\theta} - \theta^{T} \gamma_{9} D \theta ) - 
{1\over g_{s}} \theta^{T} \gamma_{i} [X^{i},\theta] ) . 
\eea

This action is second quantized in the sense that it describes multiple 
interacting strings. One can further consider the special case of free 
strings, recovered in the infra--red limit with $g_{s}=0$. In this 
limit the above two dimensional gauge theory becomes strongly 
coupled -- as the Yang--Mills gauge coupling is related to the string 
coupling as $g_{YM} \sim {1 \over g_{s}}$ -- and a non--trivial 
conformal field theory will describe the IR fixed point 
\cite{Dijkgraaf-Verlinde-Verlinde}. One can observe that in this 
limit, $g_{s} \to 0$, the world--sheet gauge field drops out, and 
moreover all matrices are diagonalized, {\it i.e.}, they will commute,

\be
[X^{i},X^{j}]=0, \ \ \ \ \ \ \ \ [X^{i},\theta]=0 . \nonumber
\ee

\noindent
In this conformal field theory limit the previous Matrix string 
action reduces to,

\be
S = {1\over 2\pi} \int d\tau d\sigma \left( {1\over 2} 
(\partial_{\mu} X^{i})^{2} + i \theta^{T} \rho^{\mu} 
\partial_{\mu} \theta \right),
\ee

\noindent
where $\{ \mu \}$ are world--sheet indices. This action can be exactly 
mapped to the light--cone Green-Schwarz action for the Type II superstring 
\cite{Bonelli-Bonora-Nesti-1, Bonelli-Bonora-Nesti-2, 
Bonelli-Bonora-Nesti-Tomasiello-1, Bonelli-Bonora-Nesti-Tomasiello-2, 
Bonelli}.

We have gone through a lengthy review of the DVV reduction, in order 
to set pace and notation for the section that follows. In there, we shall 
follow the same procedure applied to the full set of multipole moments of the 
11--dimensional supercurrents for the stress tensor $T^{IJ}$, membrane
current $J^{IJK}$ and fivebrane current $M^{IJKLMN}$. These ``DVV 
reduced'' tensors will be the basis for the matrix string theory 
action in a weakly curved background.

\section{Reduction via the 9--11 Flip}

In order to write down the matrix string theory action in weak 
background fields, one needs to know the DVV reduction of the Matrix 
theory stress tensor, membrane current and 5--brane current. This 
should be clear from the fact that the matrix string theory action is 
obtained via a DVV reduction of the Matrix theory action (as explained 
in the previous section), and the fact that in weak background fields 
the Matrix theory action is constructed precisely with the use of 
these tensors and currents (as explained in section 2, in 
particular in expression (\ref{mta})). We will begin in here by 
applying the DVV reduction using the $9-11$ flip, just as described 
previously. Later, we shall look at the sequence of dualities, and compare
both procedures.

Let us start by specifying the conventions for the following of this section. 
Time derivatives are taken with respect to Minkowski time $t$. All 
expressions have been written in a gauge with $A_0 = 0$.  Gauge
invariance can be restored by replacing $\dot{X}$ with $D_t X$.
Indices $i,j,\ldots$, run from 1 through 9, while indices $a,b,
\ldots$, run from 0 through 9. In these expressions we use the 
definitions $F_{0i} = \dot{X}^{i},\; F_{ij} = i[X^{i}, X^{j}]$. A 
Matrix form for the transverse 5--brane current components 
$M^{+ijklm},M^{ijklmn}$ is as yet unknown. There are also fermionic 
components of the supercurrent which couple to background fermion fields 
in the supergravity theory.  We will not discuss these couplings in this 
paper, but the Matrix theory form of the currents is determined in 
\cite{Mark-Wati-3}. Moreover, there is also a 6--brane current appearing 
in Matrix theory related to nontrivial 11--dimensional background metrics.

\subsection{Matrix Theory Tensors}

We shall briefly describe the DVV reduction of the first component of 
the stress tensor, referring the specifics to the full description 
in the previous section. Then, we shall simply present the results for 
the other components in a schematic form (in the Appendix).

The zeroth moment of the $T^{++}$ component of the Matrix stress 
tensor is given by,


\be
T^{++} = {1 \over R}\str\left(\identity\right),
\ee

\noindent
where \str indicates a trace which is symmetrized over all orderings 
of terms of the forms $F_{ab}$, $\theta$ and $[X^{i},\theta]$. We 
shall denote by the same name, $T^{++}$, the time integrated component 
which appears in the curved Matrix theory action. It is to this 
integrated term that we will apply the DVV reduction. In this term 
there is no need to introduce explicit dimensionfull parameters -- 
there are no background fields -- but we shall do it automatically in 
all the following terms, just as we did for the Matrix theory 
action in the previous section. As the theory is further compactified 
along the $9^{th}$ direction, we $T$--dualize to obtain, after the 
$9-11$ flip,

\be
T^{++} = \int dt\; {1 \over R}{g_{s} \over 2\pi\ell_{s}} 
\int_{0}^{2\pi{\ell_{s}\over g_{s}}} d\hat{x}\; 
\str\left(\identity\right). \nonumber
\ee

\noindent
Rescaling of world--sheet coordinates, background fields, and coupling 
constants (most of them trivial for this component), we are left with 
the final result,


\be
T^{++} = {1 \over 2\pi} \left({\ell_{s} \over R}\right)^{2} 
\int d\sigma d\tau\; \str\left(\identity\right).
\ee

\noindent
Moreover, we shall later be interested in the conformal field theory 
limit of these tensors. So, we further observe that the free string 
limit can be easily taken as,


\be
\lim_{g_{s}\to 0} T^{++} = T^{++}.
\ee

We can proceed along the same line for the following components. The 
zeroth moment of the $T^{+i}$ component of the Matrix stress tensor 
is given by,


\be
T^{+i} = {1 \over R}\str\left(\dot{X_i}\right).
\ee

\noindent
Under $T$--duality for the $9-11$ flip, one obtains for $i \neq 9$,

\be
T^{+i} = \int dt\; {1 \over R}{g_{s} \over 2\pi\ell_{s}} 
\int_{0}^{2\pi{\ell_{s}\over g_{s}}} d\hat{x}\; 
\str\left(\dot{X_i}\right). \nonumber
\ee

\noindent
After the needed rescalings of world--sheet coordinates, background 
fields, and coupling constants, we are left with the final result,


\be
T^{+i} = {1 \over 2\pi} \left({\ell_{s} \over R}\right) 
\int d\sigma d\tau\; \str\left(\dot{X_i}\right).
\ee

\noindent
As to the free string limit, it can be taken as,


\be
\lim_{g_{s}\to 0} T^{+i} = T^{+i}.
\ee

\noindent
Under $T$--duality for the $9-11$ flip, one obtains for $i=9$,

\be
T^{+9} = \int dt\; {1 \over R}{g_{s} \over \ell_{s}} 
\int_{0}^{2\pi{\ell_{s}\over g_{s}}} d\hat{x}\; 
\str\left(\ell_{s}^{2} \dot{A}\right). \nonumber
\ee

\noindent
After the needed rescalings of world--sheet coordinates, background 
fields, and coupling constants, we are left with the final result,


\be
T^{+9} = {1 \over 2\pi} \left({\ell_{s} \over R}\right) 
\int d\sigma d\tau\; \str\left(g_{s} \dot{A}\right).
\ee

\noindent
As to the free string limit, it can be taken as,


\be
\lim_{g_{s}\to 0} T^{+9} = 0.
\ee

The procedure is always the same, for all the components. It should 
be clear to the reader how to obtain all the results, which are 
presented schematically in the Appendix. A few comments can be made, about 
the structures we have derived. First, as was trivially expected, the string 
coupling appears as expected, {\it i.e.}, a factor of $g_{s}$ for each factor 
of $\dot{A}$, and factor of ${1\over g_{s}}$ for each factor of 
$[X,X]$, or for each factor of $[X,\theta]$. Moreover, every tensor 
(and the action in section 3, also) has an overall normalization 
factor of ${1\over 2\pi}$. Second, and more 
importantly, we observe that if one counts operator insertions of 
background coordinates into the currents as $\dot{X}$, $\dot{A}$, 
$\theta$, $[X,X]$ and $[X,\theta]$ each counting as one operator 
insertion, then the order of the currents depends on the number of 
insertions as follows. For zero insertions, it is order 
${\cal O}=({\ell_{s} \over R})^{2}$; for one insertion, it is order 
${\cal O}=({\ell_{s} \over R})$; for two insertions, it is order 
${\cal O}=1$; for three insertions, it is order 
${\cal O}=({R\over \ell_{s}})$; for four insertions, it is order 
${\cal O}=({R\over \ell_{s}})^{2}$; and so on, for $n$ insertions it 
is order ${\cal O}=({R\over \ell_{s}})^{n-2}$.

\subsection{Matrix String Theory Tensors}

We have thus performed the analysis of the Matrix theory expressions 
for the stress tensor, the membrane current and the 5--brane current. 
As previously explained in section 2, one can obtain the matrix string 
theory action in terms of other tensors: the sources $I_{p}$ of 
$Dp$--brane currents for $p=2n$, the sources $I_{s}$ and $I_{5}$ 
associated with fundamental string and $NS5$--brane currents respectively, 
and also the sources $I_{h}$ and $I_{\phi}$ of background metric and 
background dilaton fields. These currents $I$ can moreover be expressed 
as linear combinations of the Matrix theory expressions for $T$, $J$ and 
$M$. In previous work, the results for the lowest dimension operators 
appearing in the monopole (integrated) $D0$--brane currents were 
obtained \cite{Mark-Wati-3,Mark-Wati-4,Mark-Wati-5}.

Now, because of the $9-11$ flip, we are dealing with $M$--theory on 
spacelike $R_{9}$ instead of $M$--theory on spacelike $R_{11}$ as 
before. This means that the $I$ tensors are not necessarily related to 
the $T$, $J$ and $M$ tensors in the same way as in the case of the 
$D0$--brane action that was described briefly in section 2. We begin 
by addressing such a question, in order to derive the correct expressions 
for the $I$ linear tensor couplings. The original $M$--theory where the 
$D0$--brane couplings were derived was spacelike compactified along 
$R_{11}$, so that in light--cone coordinates we would be dealing 
with $X^{\pm} \sim X^{0} \pm X^{11}$ and a further compact 
coordinate $X^{9}$. With the $9-11$ flip we are now led to an 
$\hat{M}$--theory compactified along $R_{9}$, and where in light--cone 
gauge the coordinates are now $\hat{X}^{\pm} \sim \hat{X}^{0} \pm 
\hat{X}^{9}$ and the compact coordinate $\hat{X}^{11}$. Clearly we 
have two frames, the ``11'' frame in the original $M$--theory, and the 
``9'' frame in the flipped $\hat{M}$--theory. The relations we 
presented briefly in section 2 concerning the relation between the 
$I$ tensors and the Matrix theory tensors $T$, $J$ and $M$, are still 
valid in the flipped ``9'' frame, but now relating the $I$ tensors to 
the Matrix tensors in this frame, {\it i.e.}, $\hat{T}$, $\hat{J}$ and 
$\hat{M}$. If we then relate these ``9'' frame Matrix tensors back to 
the ``11'' frame Matrix tensors, we will be able to express the matrix 
string theory couplings $I$ in terms of the just derived DVV reduced 
Matrix tensors $T$, $J$ and $M$. So, all one needs to do is a simple 
change of coordinates.

To begin with a simple example, let us look at the $I^{ij}_{s}$ 
component of the matrix fundamental string current, which is given by:

\be
I^{ij}_{s} \sim \hat{J}^{+ij} - \hat{J}^{-ij}.
\ee

\noindent
This expression holds in the ``9'' frame. Relating the $\hat{J}$ 
tensor to the $J$ tensor in the ``11'' frame, one obtains,

\be
I^{ij}_{s} \sim \hat{J}^{+ij} - \hat{J}^{-ij} \sim J^{9ij}.
\ee

\noindent
On the other hand, at the level of background fields, one knows how to 
relate the NS 2--form $B_{\mu\nu}$ to the $M$--theory 3--form $A_{IJK}$, 
via $B_{ij} \sim A_{9ij}$, where 9 is the spacelike compact direction 
involved in the Sen--Seiberg limit. The coupling we have just derived 
above is then precisely what one would expect.

The procedure is always the same, and it should be straightforward for 
the reader to reproduce the results, which we now present 
schematically. Observe that as we change from the ``9'' frame to the 
``11'' frame, there is a mixing of different orders in the currents, 
{\it i.e.}, there will be tensors in different powers of ${\cal O} 
\left( ({R\over \ell_{s}})^{n-2} \right)$. We will neglect some of 
these tensors in the following expressions, and only keep the orders 
of interest to us. The components of the matrix string current associated 
with the background metric field are:


\bea
I^{00}_{h} &=& \hat{T}^{++} + \hat{T}^{+-} + (I^{00}_{h})_{8} + 
{\cal O}(\hat{X}^{12}) = T^{++} + T^{+-} + \cdots
\nonumber \\
&=& {1 \over 2\pi} \int d\sigma d\tau\; \str ( 
\left({\ell_{s} \over R}\right)^{2} \identity + {1\over 2}
\dot{X_{i}}^{2} + {1\over 2}(DX^{i})^{2} + {1\over 2}
g_{s}^{2}\dot{A}^{2} - {1\over 2g_{s}^{2}} \sum_{i<j} 
[X^{i},X^{j}]^{2} \nonumber \\
&
& + 
{1\over g_{s}}\theta \gamma_{i}[X^{i},\theta] + 
i\theta \gamma^{9} D\theta ) + \ldots,
\nonumber \\
I^{0i}_{h} &=& \hat{T}^{+i} + \hat{T}^{-i} + {\cal O}(\hat{X}^{10}) = 
T^{+i} + T^{-i} + \cdots
\nonumber \\
&=& {1 \over 2\pi} \int d\sigma d\tau\; \str \{
\left({\ell_{s} \over R}\right) \dot{X_i} + 
\left({R \over \ell_{s}}\right) [
{1\over 2}\dot{X_i} (\dot{X_j})^{2} + {1\over 2} g_{s}^{2} 
\dot{X_{i}}\dot{A}^{2} - {1\over 2 g_{s}^{2}} \dot{X_i} \sum_{j<k} 
[X^{j},X^{k}]^{2} \nonumber \\
&
& + 
{1\over 2} \dot{X_{i}}(D X^{j})^{2} - 
{1\over g_{2}^{2}} [X^{i},X^{j}][X^{j},X^{k}] \dot{X_k} - 
D X^{i} D X^{k} \dot{X_k} + i [X^{i},X^{j}] D X^{j} \dot{A} 
\nonumber \\ 
& 
& - 
{1\over 2g_{s}} \theta_\alpha 
\dot{X_k}[X_j,\theta_\beta] \{\gamma^k\delta_{ij} 
+\gamma^i\delta_{jk} -2\gamma^j\delta_{ki} \}_{\alpha \beta} - 
{1\over 2} \dot{A} \theta \gamma^{9} [X_i,\theta] + 
i \dot{X_i} \theta \gamma^{9} D \theta 
\nonumber\\ 
& 
& - 
{i \over 2} g_{s} \dot{A} \theta \gamma^{i} D \theta - 
{i \over 4g_{s}^{2}} \theta_{\alpha} [X^{k},X^{j}] [X^{l},\theta_{\beta}] 
\{ \gamma^{[ikjl]} + 2 \gamma^{[jl]} \delta_{ki} + 
4\delta_{ki}\delta_{jl} \}_{\alpha \beta} 
\nonumber\\ 
&  
& - 
{1 \over 2g_{s}} \theta_{\alpha} D X^{k} [X^{j},\theta_{\beta}] 
\{ \gamma^{[ik9j]} + \gamma^{[9j]} \delta_{ki} \}_{\alpha \beta} + 
{1 \over 4g_{s}} \theta_{\alpha} [X^{k},X^{j}] D \theta_{\beta} 
\{ \gamma^{[ikj9]} + 2 \gamma^{[j9]} \delta_{ki} \}_{\alpha \beta} 
\nonumber \\ 
& 
& - 
i D X^{i} \theta D \theta + \cdots ]\, \} + \dots,
\nonumber \\
I^{ij}_{h} &=& \hat{T}^{ij} + (I^{ij}_{h})_{8} + {\cal 
O}(\hat{X}^{12}) = T^{ij} + \cdots
\nonumber \\
&=& {1\over 2\pi} \int d\sigma d\tau\; \str ( \dot{X_i}\dot{X_j} 
- D X^{i} D X^{j} - {1\over g_{s}^{2}} [X^{i},X^{k}][X^{k},X^{j}] 
\nonumber \\
&
& - 
{1\over 2 g_{s}} \theta\gamma^i[X_j,\theta] - 
{1\over 2 g_{s}} \theta\gamma^j[X_i,\theta] ) + \ldots,
\eea

\noindent
where $(I^{00}_{h})_{8} = {3\over2} \hat{T}^{--} + \cdots$ and 
$(I^{ij}_{h})_{8} = 2 \hat{T}^{--} + \cdots$, and we know the matrix 
string form of $\hat{T}^{--}$ from (\ref{mst--}). The conformal 
field theory limit of these tensors is simply:


\bea \label{cftih}
\lim_{g_{s} \to 0} I^{00}_{h} &=&  {1 \over 2\pi} \int d\sigma d\tau\; 
\str \left( \left({\ell_{s} \over R}\right)^{2} \identity + 
{1\over 2}\dot{X_{i}}^{2} + {1\over 2}(\partial X^{i})^{2} + 
i\theta \gamma^{9}\partial \theta \right) + \ldots,
\nonumber \\
\lim_{g_{s} \to 0} I^{0i}_{h} &=& {1 \over 2\pi} \int d\sigma d\tau\; 
\str ( \left({\ell_{s} \over R}\right) \dot{X_i} + 
\left({R \over \ell_{s}}\right) \{
{1\over 2}\dot{X_i} (\dot{X_j})^{2} + {1\over 2} \dot{X_{i}}(\partial 
X^{j})^{2} - \partial X^{i} \partial X^{k} \dot{X_k} 
\nonumber \\ 
& 
& +
i \dot{X_i} \theta \gamma^{9} \partial \theta - 
i \partial X^{i} \theta \partial \theta \} ) + \ldots,
\nonumber \\
\lim_{g_{s} \to 0} I^{ij}_{h} &=& {1\over 2\pi} \int d\sigma d\tau\; \str 
\left( \dot{X_i}\dot{X_j} - \partial X^{i} \partial X^{j}\right) + 
\ldots.
\eea

The $I_{\phi}$ matrix string current associated to the dilaton is 
given by,


\bea \label{msc-d}
I_{\phi} &=& \hat{T}^{++} - {1\over 3} ( \hat{T}^{+-} + \hat{T}^{ii} ) + 
(I_{\phi})_{8} + {\cal O}(\hat{X}^{12}) = T^{++} + {1\over3} (T^{+-} + 
T^{ii}) + \cdots
\nonumber \\
&=& {1 \over 2\pi} \int d\sigma d\tau\; \str \left( 
\left({\ell_{s} \over R}\right)^{2} \identity 
+ {1\over 2} \dot{X_{i}}^{2} - {1\over 2}(DX^{i})^{2} + {1\over 2} 
g_{s}^{2}\dot{A}^{2} + {1\over 2 g_{s}^{2}} \sum_{i<j} [X^{i},X^{j}]^{2} 
\right) 
\nonumber \\
&
& 
+ \ldots,
\eea

\noindent
where  $(I_{\phi})_{8} = -{1\over 2} \hat{T}^{--} + \cdots$ and we 
know the matrix string form of $\hat{T}^{--}$ from (\ref{mst--}). 
This current has the following conformal field theory limit,


\be
\lim_{g_{s} \to 0} I_{\phi} = {1 \over 2\pi} 
\int d\sigma d\tau\; \str \left( \left({\ell_{s} \over R}\right)^{2} 
\identity + {1\over 2} \dot{X_{i}}^{2} - {1\over 2} (\partial 
X^{i})^{2} \right) + \ldots.
\ee

The components of the matrix fundamental string current are:


\bea
I^{0i}_{s} &=& 3 \hat{J}^{+-i} + {\cal O}(\hat{X}^{8}) = 3 J^{+i9} + 
3 J^{-i9} + \cdots
\nonumber \\
&=& {1\over 2\pi} \int d\sigma d\tau\; \str (
- {1\over 2} \left({\ell_{s}\over R}\right) D X^{i} + 
\left({R \over \ell_{s}} \right) \{
{1\over 2} \dot{X^{i}} \dot{X^{k}} D X^{k} - 
{i \over 2} \dot{A} \dot{X^{k}} [X^{k},X^{i}] + 
{g_{s}^{2} \over 4} \dot{A}^{2} D X^{i} 
\nonumber \\ 
& 
& - 
{1\over 4} (\dot{X^{k}})^{2} D X^{i} - 
{1\over 4 g_{s}^{2}} D X^{i} \sum_{k<l} [X^{k},X^{l}]^{2} - 
{1\over 4} D X^{i} (D X^{k})^{2} 
\nonumber \\ 
& 
& - 
{1\over 2 g_{s}^{2}} [X^{i},X^{k}] [X^{k},X^{l}] D X^{l} + 
{1\over 4 g_{s}} \theta_{\alpha} 
\dot{X^{k}}[X^{m},\theta_{\beta}] \{ \gamma^{[ki9m]} + 
\gamma^{[9m]} \delta_{ki} \}_{\alpha \beta}
\nonumber \\ 
& 
& - 
{1\over 4} \dot{A} \theta_{\alpha} 
[X^{m},\theta_{\beta}] \{ \gamma^{[im]} + 2 \delta_{im} \}_{\alpha 
\beta} + 
{i \over 2} \dot{X^{i}} \theta D \theta - 
{i \over 4} \dot{A} \theta \gamma^{[i9]} D \theta 
\nonumber \\ 
& 
& + 
{3i \over 4 g_{s}^{2}} \theta_{\alpha} 
[X^{k},X^{l}][X^{m},\theta_{\beta}] \{\gamma^{[9kl]} \delta_{mi} + 
2 \gamma^{[li9]} \delta_{km} + 2 \gamma^{9} \delta_{il} \delta_{km} 
\}_{\alpha \beta}
\nonumber\\
& 
& - 
{3\over 2 g_{s}} D X^{k} \theta \gamma^{k} [X^{i},\theta] - 
{3\over 2 g_{s}} D X^{k} \theta \gamma^{i} [X^{k},\theta] 
\nonumber \\ 
& 
& + 
{3i \over 4 g_{s}} \theta_{\alpha} 
[X^{k},X^{l}] D \theta_{\beta} \{ \gamma^{[ikl]} + 2 
\gamma^{l} \delta_{ik} \}_{\alpha \beta} - 
i D X^{i}\theta \gamma^{9} D \theta + \cdots \} ) + \ldots,
\nonumber \\
I^{ij}_{s} &=& 3 \hat{J}^{+ij} - 3 \hat{J}^{-ij} + {\cal O}(\hat{X}^{10}) = 
3 J^{9ij} + \cdots
\nonumber \\
&=& {1\over 2\pi} \int d\sigma d\tau\; \str\left(
- {1\over 2} \dot{X^{i}}D X^{j} + {1\over 2} \dot{X^{j}}D X^{i} - 
{i \over 2} \dot{A}[X^{i},X^{j}] + 
{1\over 4 g_{s}} \theta \gamma^{[ij9l]}[X_l,\theta] \right).
\eea

\noindent
The conformal field theory limit of these tensors is:


\bea \label{cftis}
\lim_{g_{s} \to 0} I^{0i}_{s} &=& 
{1\over 2\pi} \int d\sigma d\tau\; \str (
- {1\over 2} \left({\ell_{s}\over R}\right) \partial X^{i} + 
\left({R \over \ell_{s}} \right) \{
{1\over 2} \dot{X^{i}} \dot{X^{k}} \partial X^{k} 
\nonumber \\ 
& 
& - 
{1\over 4} (\dot{X^{k}})^{2} \partial X^{i} - 
{1\over 4} \partial X^{i} (\partial X^{k})^{2} + 
{i \over 2} \dot{X^{i}} \theta \partial \theta - 
i \partial X^{i}\theta \gamma^{9} \partial \theta\} ) + \ldots,
\nonumber \\
\lim_{g_{s} \to 0} I^{ij}_{s} &=& 
{1\over 2\pi} \int d\sigma d\tau\; \str\left(
- {1\over 2} \dot{X^{i}} \partial X^{j} + {1\over 2} \dot{X^{j}} 
\partial X^{i} \right) + \ldots.
\eea

Let us now move to the R--R fields. The components of the matrix string 
$D0$--brane current are:


\bea
I^{0}_{0} &=& \hat{T}^{++} = T^{+9} + T^{-9} + \cdots 
\nonumber \\
&=& {1 \over 2\pi} \int d\sigma d\tau\; \str ( 
\left({\ell_{s} \over R}\right) g_{s} \dot{A}
+ \left({R \over \ell_{s}}\right) \{
{1\over 2} g_{s}\dot{A} (\dot{X^{i}})^{2} + {1\over 2} g_{s}^{3}\dot{A}^{3} 
- {1\over 2 g_{s}} \dot{A} \sum_{i<j} [X^{i},X^{j}]^{2} 
\nonumber \\
&
& - 
{i \over g_{s}} D X^{i} [X^{i},X^{j}] \dot{X^{j}} - 
{1\over 2} g_{s} (D X^{i})^{2} \dot{A} - 
{1 \over 2 g_{s}} \dot{X^{i}} \theta 
\gamma^{9} [X^{i},\theta] + \dot{A} \theta \gamma^{i} [X^{i},\theta] 
\nonumber \\ 
& 
& - 
{i \over 2} \dot{X^{i}} \theta \gamma^{i} D\theta - {i \over 4 
g_{s}^{2}} [X^{i},X^{j}] \theta \gamma^{[9ijk]} [X^{k},\theta] + 
{1 \over 2 g_{s}} \theta_{\alpha} D X^{i} [X^{j},\theta_{\beta}] 
\{ \gamma^{[ij]} + 2\delta_{ij} \}_{\alpha \beta} 
\nonumber \\ 
& 
& + 
{i \over 2} D X^{i} \theta \gamma^{[i9]} D \theta + \cdots \} ) +\ldots,
\nonumber \\
I^{i}_{0} &=& \hat{T}^{+i} = T^{9i} + \cdots
\nonumber \\
&=& {1\over 2\pi} \int d\sigma d\tau\; \str \left( g_{s} 
\dot{X_i}\dot{A} + {i \over g_{s}} [X^{i},X^{k}] D X^{k} - 
{i \over 2} \theta\gamma^{i} D \theta - {1\over 2g_{s}} 
\theta\gamma^{9} [X_i,\theta] \right) + \ldots.
\eea

\noindent
With the conformal field theory limit:


\bea
\lim_{g_{s} \to 0} I^{0}_{0} &=& 
{1 \over 2\pi} \left({R \over \ell_{s}}\right) 
\int d\sigma d\tau\; \str \left( - 
{i \over 2} \dot{X^{i}} \theta \gamma^{i} \partial \theta + 
{i \over 2} \partial X^{i} \theta \gamma^{[i9]} \partial \theta
\right) + \ldots,
\nonumber \\
\lim_{g_{s} \to 0} I^{i}_{0} &=& 
{1\over 2\pi} \int d\sigma d\tau\; \str \left( - 
{i \over 2} \theta\gamma^{i} \partial \theta \right) + \ldots.
\eea

\noindent
Observe that these expressions for the $D0$--brane current are exact 
in the hatted frame, unlike all the other expressions for the matrix string 
theory currents, which are given up to higher order terms in the coordinate 
fields, ${\cal O}(\hat{X}^{n})$.

The components of the matrix string theory $D2$--brane current are:


\bea
I^{0ij}_{2} &=& \hat{J}^{+ij} + {\cal O}(\hat{X}^{10}) = J^{+ij} + 
J^{-ij} + \cdots
\nonumber \\
&=& {1\over 2\pi} \int d\sigma d\tau\; \str ( 
- \left({\ell_{s}\over R}\right) {i \over 6 g_{s}} [X^{i},X^{j}] 
+ \left({R \over \ell_{s}} \right) \{
{i \over 6 g_{s}} \dot{X^{i}} \dot{X^{k}} [X^{k},X^{j}] 
\nonumber \\
&
& - 
{i \over 6 g_{s}} \dot{X^{j}} \dot{X^{k}} [X^{k},X^{i}] - 
{1\over 6} g_{s} \dot{A} \dot{X^{i}} D X^{j} + 
{1\over 6} g_{s} \dot{A} \dot{X^{j}} D X^{i} 
\nonumber \\
&
& - 
{i \over 12 g_{s}} (\dot{X^{k}})^{2} [X^{i},X^{j}] - 
{i \over 12} g_{s} \dot{A}^{2} [X^{i},X^{j}] - 
{i \over 12 g_{s}^{3}} [X^{i},X^{j}] \sum_{k<l} [X^{k},X^{l}]^{2} 
\nonumber \\ 
& 
& + 
{i \over 12 g_{s}} [X^{i},X^{j}] (D X^{k})^{2} - 
{i \over 6 g_{s}^{3}} [X^{i},X^{k}] [X^{k},X^{l}] [X^{l},X^{j}] - 
{i \over 6 g_{s}} D X^{i} D X^{k} [X^{k},X^{j}] 
\nonumber \\ 
& 
& + 
{i \over 6 g_{s}} D X^{j} D X^{k} [X^{k},X^{i}] + 
{1\over 12 g_{s}} \theta_{\alpha} 
\dot{X^{k}}[X^{m},\theta_{\beta}] \{ \gamma^{[kijm]} + 
\gamma^{[jm]} \delta_{ki} - \gamma^{[im]} \delta_{kj} 
\nonumber \\ 
& 
& \hspace{1in} + 
2 \delta_{jm} \delta_{ki} - 2 \delta_{im} \delta_{kj}\}_{\alpha \beta}
\nonumber \\ 
& 
& + 
{1\over 12} \dot{A} \theta \gamma^{[9ijm]} [X^{m},\theta] + 
{i \over 12} \theta_{\alpha} \dot{X^{k}} D \theta_{\beta} \{ 
\gamma^{[kij9]} + \gamma^{[j9]} \delta_{ki} - \gamma^{[i9]} 
\delta_{kj} \}_{\alpha \beta}
\nonumber\\
& 
& + 
{i \over 4 g_{s}^{2}} \theta_{\alpha} 
[X^{k},X^{l}][X^{m},\theta_{\beta}] \{\gamma^{[jkl]} 
\delta_{mi} - \gamma^{[ikl]} \delta_{mj} + 2 \gamma^{[lij]} \delta_{km} 
+ 2 \gamma^{l} \delta_{jk} \delta_{im} 
\nonumber\\
& 
& \hspace{1in} - 
2 \gamma^{l} \delta_{ik} \delta_{jm} + 2 \gamma^{j} \delta_{il} 
\delta_{km} - 2 \gamma^{i} \delta_{jl} \delta_{km}\}_{\alpha \beta}
\nonumber\\ 
& 
& +
{1\over 2 g_{s}} \theta_{\alpha} D X^{k} [X^{m},\theta_{\beta}] 
\{ \gamma^{[jk9]} \delta_{mi} - \gamma^{[ik9]} \delta_{mj} + 
\gamma^{[9ij]} \delta_{km} + 
\gamma^{9} \delta_{jk} \delta_{im} - 
\gamma^{9} \delta_{ik} \delta_{jm} \}_{\alpha \beta} 
\nonumber \\ 
& 
& - 
{i \over 2} \theta_{\alpha} D X^{l} D \theta_{\beta} \{\gamma^{[lij]} 
+ \gamma^{j} \delta_{il} - \gamma^{i} \delta_{jl} \}_{\alpha \beta} + \cdots \} ) + \ldots,
\nonumber \\
I^{ijk}_{2} &=& \hat{J}^{ijk} + {\cal O} (\hat{X}^{8}) = J^{ijk} + \cdots
\nonumber \\
&=& {1\over 2\pi} \int d\sigma d\tau\; \str (
- {i \over 6 g_{s}} \dot{X^{i}}[X^{j},X^{k}] - 
{i \over 6 g_{s}} \dot{X^{j}}[X^{k},X^{i}] - 
{i \over 6 g_{s}} \dot{X^{k}}[X^{i},X^{j}] 
\nonumber \\ 
& 
& + 
{1\over 12 g_{s}} \theta \gamma^{[ijkl]}[X_l,\theta] + 
{i \over 12} \theta \gamma^{[ijk9]} D \theta ) 
+ \ldots.
\eea

\noindent
The conformal field theory limit for the membrane current is:


\bea
\lim_{g_{s} \to 0} I^{0ij}_{2} &=& 
{1\over 2\pi} \left({R \over \ell_{s}} \right) 
\int d\sigma d\tau\; \str (
{i \over 12} \theta_{\alpha} \dot{X^{k}} \partial \theta_{\beta} \{ 
\gamma^{[kij9]} + \gamma^{[j9]} \delta_{ki} - \gamma^{[i9]} 
\delta_{kj} \}_{\alpha \beta}
\nonumber\\
& 
& - 
{i \over 2} \theta_{\alpha} \partial X^{l} \partial \theta_{\beta} 
\{\gamma^{[lij]} + \gamma^{j} \delta_{il} - \gamma^{i} \delta_{jl} 
\}_{\alpha \beta} ) + \ldots,
\nonumber \\
\lim_{g_{s} \to 0} I^{ijk}_{2} &=& {1\over 2\pi} 
\int d\sigma d\tau\; \str\left( 
{i \over 12} \theta \gamma^{[ijk9]} \partial \theta \right) 
+ \ldots.
\eea

Moving towards the $D4$--brane current, the components are given by:


\bea
I^{0ijkl}_{4} &=& 6 \hat{M}^{+-ijkl} + {\cal O}(\hat{X}^{8}) = 6 
M^{-i9jkl} + \cdots
\nonumber \\
&=& {1\over 2\pi} \left({R \over \ell_{s}}\right) 
\int d\sigma d\tau\; \str (
- {30 i \over g_{s}}\dot{X}^{[i}[X^{j},X^{k}] D X^{l]} - 
{15\over 2 g_{s}}\dot{A} [X^{[i},X^{j}][X^{l},X^{k]}]
\nonumber \\ 
& 
& + 
5 i\theta\dot{X}^{[i}\gamma^{jkl]} D \theta - 
5 \dot{A} \theta \gamma^{[ijl} [X^{k]},\theta] - 
{15\over g_{s}} \theta\dot{X}^{[i}\gamma^{|9|kl} [X^{j]},\theta] 
\nonumber \\ 
& 
& + 
{15i \over 2 g_{s}^{2}} \theta [X^{[i},X^{j}]\gamma^{kl]9} \gamma^{n} 
[X^{n},\theta] - {15\over 2 g_{s}} \theta [X^{[i},X^{j}]\gamma^{kl]9} 
\gamma^{9} D \theta 
\nonumber \\ 
& 
& + 
{5\over g_{s}} \theta D X^{[i}\gamma^{jlk]} \gamma^{n} 
[X^{n},\theta] + 5i \theta D X^{[i}\gamma^{jlk]} 
\gamma^{9} D \theta ) + \ldots,
\nonumber \\
I^{ijklm}_{4} &=& - 6 \hat{M}^{-ijklm} + {\cal O}(\hat{X}^{10}) = - 6 
M^{-ijklm} + \cdots
\nonumber \\
&=& {1\over 2\pi} \left({R \over \ell_{s}}\right) 
\int d\sigma d\tau\; \str (
{15\over 2 g_{s}^{2}}\dot{X}^{[i}[X^{j},X^{k}][X^{l},X^{m]}]
+ {5\over g_{s}}\theta\dot{X}^{[i}\gamma^{jkl}[X^{m]},\theta] 
\nonumber \\ 
& 
& + 
{5i \over 2 g_{s}^{2}}
\theta [X^{[i},X^{j}]\gamma^{klm]}\gamma^{n} [X^{n},\theta] - 
{5\over 2 g_{s}}
\theta [X^{[i},X^{j}]\gamma^{klm]} \gamma^{9} D \theta 
) + \ldots.
\eea

\noindent
The conformal field theory limit for the 4--brane current is:


\bea
\lim_{g_{s} \to 0} I^{0ijkl}_{4} &=& {1\over 2\pi} 
\left({R \over \ell_{s}}\right) \int d\sigma d\tau\; \str 5i \left(
\theta\dot{X}^{[i}\gamma^{jkl]} \partial \theta 
+ \theta \partial X^{[i}\gamma^{jlk]} \gamma^{9} \partial \theta 
\right) + \ldots,
\nonumber \\
\lim_{g_{s} \to 0} I^{ijklm}_{4} &=& {\cal O}(X^{10}) + \ldots.
\eea

Next we analyze the $D6$--brane current in matrix string theory. The 
components of this current are given by:


\bea
I^{0ijklmn}_{6} &=& \hat{S}^{+ijklmn} + {\cal O}(\hat{X}^{10}) = 
S^{+ijklmn} + \cdots
\nonumber \\
&=& {1\over 2\pi} \left({R \over \ell_{s}}\right) 
\int d\sigma d\tau\; \str\left(- {i \over g_{s}^{3}} 
[X^{[i},X^{j}][X^{k},X^{l}][X^{m},X^{n]}] \right) 
+ \ldots,
\nonumber \\
I^{ijklmnp}_{6} &=& \hat{S}^{ijklmnp} + {\cal O}(\hat{X}^{12}) = 
S^{ijklmnp} + \cdots
\nonumber \\
&=& {1\over 2\pi} \left({R \over \ell_{s}}\right)^{2} 
\int d\sigma d\tau\; \str\left(
-{7i \over g_{s}^{3}} 
[X^{[i},X^{j}][X^{k},X^{l}][X^{m},X^{n}]\dot{X}^{p]} + 
{\cal O}(\theta^{2},\theta^{4}) \right) 
\nonumber \\ 
&
& + \ldots.
\eea

\noindent
As to the conformal field theory limit for the 6--brane current, it is:


\bea
\lim_{g_{s} \to 0} I^{0ijklmn}_{6} &=& {\cal O}(X^{10}) + \ldots,
\nonumber \\
\lim_{g_{s} \to 0} I^{ijklmnp}_{6} &=& {\cal O}(X^{12}) + \ldots.
\eea

A Matrix theory form for the transverse $M5$--brane current components 
$M^{+ijklm}$ and $M^{ijklmn}$ is not known (though it is believed that 
these operators identically vanish in this light--front gauge). 
Therefore, we cannot determine the $NS5$--brane current components 
$I^{ijklmn}_{5}$ and $I^{0ijklm}_{5}$ (though most likely these 
operators will vanish in the Type IIA description as well, at least to 
the lowest order we are considering here). For further discussions on 
these points, see \cite{Mark-Wati-4}.

With these matrix string currents, the sigma model action for matrix string 
theory in weakly curved backgrounds is then simply written as:


\bea \label{mstawcb}
{\cal S} &=& {1 \over 2\pi} \int d\sigma d\tau\; ( 
{1\over 2} g_{\mu\nu}^{IIA} (X) I^{\mu\nu}_{g} + \phi (X) I_{\phi} + 
B_{\mu\nu} (X) I^{\mu\nu}_{s} + \widetilde{B}_{\mu\nu\lambda\rho\sigma\tau} (X) 
I^{\mu\nu\lambda\rho\sigma\tau}_{5} 
\nonumber \\
&
& + 
C_{\mu} (X) I^{\mu}_{0} + \widetilde{C}_{\mu\nu\lambda\rho\sigma\tau\xi} (X) 
I^{\mu\nu\lambda\rho\sigma\tau\xi}_{6} + C_{\mu\nu\lambda} (X) 
I^{\mu\nu\lambda}_{2} + \widetilde{C}_{\mu\nu\lambda\rho\sigma} (X) 
I^{\mu\nu\lambda\rho\sigma}_{4} ),
\eea

\noindent
where the notation is as follows. The metric is $g_{\mu\nu} = 
\eta_{\mu\nu} + h_{\mu\nu}$, so that the first term includes 
naturally, to linear order in $h_{\mu\nu}$, the term relative to the 
matrix string action in flat space and the linear coupling term 
$h_{\mu\nu}(X) I^{\mu\nu}_{h}$ previously derived. Also, we have seen 
that the currents $I$ that we derived were integrated currents, 
including an explicit world--sheet integration. In (\ref{mstawcb}) 
this world--sheet integration, as well as the ${1\over 2\pi}$ factor, 
have been brought out of the expressions for the current, in order to 
stress that we have obtained a two dimensional matrix gauged sigma model 
field theory. Finally, recall from 
\cite{Mark-Wati-3,Mark-Wati-4,Mark-Wati-5,Myers} what is the prescription to 
include explicit spacetime dependence in all the NS--NS and R--R 
fields. We should take the following definition:

\be
\phi(X) I_{\phi} \equiv \sum^{\infty}_{n=0} {1 \over n!} 
(\partial_{k_{1}} \cdots \partial_{k_{n}} \phi) (0) I^{(k_{1} \cdots 
k_{n})}_{\phi},
\ee

\noindent
where $I^{(k_{1} \cdots k_{n})}_{\phi}$ are the higher moments of the 
matrix string current for the dilaton. Similarly for all the other 
fields. Observe that in this work we have just analyzed zeroth moments of 
the Matrix and matrix string currents. Moreover, we shall assume for the 
remainder of the paper that the background fields satisfy the source--free 
equations of motion of Type IIA supergravity. In this case, the dual 
fields $\widetilde{C}^{D6}$, $\widetilde{B}^{NS5}$ and 
$\widetilde{C}^{D4}$ are well defined $(p+1)$--form fields given at 
linear order by,

\be
d \widetilde{C}^{D6} = \star \; d C^{D0}, \;\;\;\;
d \widetilde{B}^{NS5} = \star \; d B, \;\;\;\;
d \widetilde{C}^{D4} = \star \; d C^{D2}.
\ee

In one sentence, the tensors we have just derived allow us to build a 
matrix sigma model for the IIA string. Recall from the standard sigma 
model approach that the background fields are an infinite number of 
couplings from the point of view of the world--sheet quantum field 
theory. If the target space has characteristic curvature $R$, then 
derivatives of the metric will be of order ${1\over R}$, and so the 
effective dimensionless coupling in the theory will be 
${{\sqrt{\alpha'}}\over R} = {\ell_{s} \over R}$, quite similar to 
what we have obtained (even though previously $R$ was simply the 
radius of the compact dimension). For $R \gg \ell_{s}$ the effective 
coupling is small and perturbation theory on the world--sheet is 
useful. In this regime, the string is effectively point--like, and 
one can also use the low energy effective field theory to deal with 
the problem. To these results there will naturally be stringy 
corrections. They can be obtained from the multi--loop corrections to 
the world--sheet beta functions, as a power series in our effective 
coupling ${\ell_{s} \over R}$.

One expects that a similar story should take place in the case of the 
matrix sigma model we have derived, that describes multiple string 
interactions in non--trivial background fields. Indeed, it would be 
quite interesting to derive matrix beta functions, and therefore 
matrix equations of motion for the background fields. While at the 
level of Matrix theory one could expect that a large $N$ 
renormalization group analysis should be required \cite{Douglas-talk}, 
here at the level of matrix string theory it should only be a direct 
generalization of the one string sigma model field theory.

\section{Reduction via the $T$--$S$--$T$ Duality Sequence}

If one recalls section 3, performing the DVV reduction via the $9-11$ 
flip should be equivalent to a specific set of dualities, namely 
$T$--duality from IIA to IIB, $S$--duality of IIB, and then 
$T$--duality back to the IIA theory. We would now like to explicitly 
check such a procedure in the presence of non--trivial backgrounds. 
This would amount to a further check of the internal structure of 
string dualities. Basically, all one needs to do is DVV reduce according 
to the sequence of dualities as applied to the background fields and 
to the world--volume fields, and observe that one will obtain the 
same result as in the previous section. The transformation of the 
background fields under $T$ and $S$ dualities is well known. As to the 
transformation of the world--volume fields, it is well known for the 
case of $T$--duality as discussed for the general case in 
\cite{WT-compact}. For $S$--duality, it is not known how the world--volume 
fields transform, as we are dealing with a nonabelian gauge theory. For 
the case of the $D3$--brane, this has been discussed recently in 
\cite{Mark-Wati-5}. In here, we shall be dealing with a $D1$--brane. 
We will obtain the $S$--duality transformations for the world--volume 
fields of the 2--dimensional gauge theory by demanding consistency with 
the whole structure of matrix string theory. Also, we shall work this out 
explicitly only for a couple of terms, not for the whole series of 
components of the several matrix string supercurrents. Moreover, we 
will neglect all terms involving fermions throughout this section. 
From the previous section we already known how they appear in the tensor 
structures that compose matrix string theory, so that we can always 
take them from there when they are required in the following sections. 
However, for the purpose of checking the duality sequence it should be 
enough to look at the bosonic part of the action alone. Extending our 
results to all the components and including fermions should present no 
obvious obstacles.

We begin by following closely \cite{Mark-Wati-5}, in particular their 
discussion of implementing $T$ and $S$ dualities for the linear 
supergravity backgrounds, and somewhat the implementation of 
$T$--duality at the world--volume level. Then, due to the matching 
between the duality sequence and the $9-11$ flip, we determine the 
$S$--duality transformation rules at the level of the 2--dimensional 
world--volume theory.

Let us focus on $T$--duality, and how it acts both on the 
supergravity background fields and on the fields that live on the 
$D$--brane. Recall that when we begin the sequence of dualities we are 
looking at the world--volume theory of $D0$--branes. $T$--duality 
acting on arbitrary backgrounds independent of the compact directions 
is well known in string theory \cite{Bergshoeff-Hull-Ortin}. The 
standard $T$--duality rules can be linearized and these are the 
transformations we shall be interested in, given that we are working 
in weak background fields. Using barred indices 
$\bar{\alpha},\bar{\beta},\ldots$, for the compact directions in which 
a $T$--duality is performed and indices $\mu,\nu,\ldots$, for the 
remaining $10-p$ spacetime dimensions (including 0), we can write the 
action of $T$--duality in the linear supergravity background fields 
as \cite{Mark-Wati-5},

\bea
h_{\mu\nu} &\longleftrightarrow& h_{\mu\nu}
\nonumber \\
B_{\mu\nu} &\longleftrightarrow& B_{\mu\nu}
\nonumber \\
h_{\mu\bar{\alpha}} &\longleftrightarrow& -B_{\mu\bar{\alpha}}
\nonumber \\
h_{\bar{\alpha}\mu} &\longleftrightarrow& B_{\bar{\alpha}\mu}
\nonumber \\
h_{\bar{\alpha}\bar{\beta}} &\longleftrightarrow& -h_{\bar{\alpha}\bar{\beta}}
\nonumber \\
B_{\bar{\alpha}\bar{\beta}} &\longleftrightarrow& -B_{\bar{\alpha}\bar{\beta}}
\nonumber \\
\phi &\longleftrightarrow& \phi -{1\over2} \sum_{\bar{\alpha}} 
h_{\bar{\alpha}\bar{\alpha}}
\nonumber \\
C^{(q)}_{\mu_{1}\cdots\mu_{q-k}\bar{\alpha}_{1}\cdots\bar{\alpha}_{k}} 
&\longleftrightarrow& {1\over (n-k)!} 
\epsilon^{\bar{\alpha}_{1}\cdots\bar{\alpha}_{k}} 
C^{(q-2k+n)}_{\mu_{1}\cdots\mu_{q-k}\bar{\alpha}_{k+1}\cdots\bar{\alpha}_{n}}
\eea

\noindent
where the $(q)$ superscript indicates the $q$--form R--R field 
associated to a $D(q-1)$--brane.

The implementation being clear at the background level, let us look 
at the world--volume level. The low energy effective field theory 
living on the world--volume of a $Dp$--brane in flat background space 
is the dimensional reduction of 10--dimensional SYM theory to the 
$p+1$ world--volume dimensions. One thing one can do 
\cite{Mark-Wati-5} is to retain 10--dimensional notation for all the 
$Dp$--brane world--volume theories and reinterpret the resulting 
expressions appropriately for each case, {\it i.e.}, if we choose 
$a,b,\ldots$, as world--volume indices and $i,j,\ldots$, as indices 
transverse to the brane, then we would reinterpret expressions such 
as $F_{ai} \equiv D_{a} X^{i}$ and $F_{ij} \equiv i[X^{i},X^{j}]$. We 
therefore see that the action of $T$--duality on expressions which have 
been written in terms of the 10--dimensional notation simply amounts to 
an adequate reintrepertation of such a notation. There is only one point 
one should take into account, namely we should be careful when considering 
transverse fields $X^{i}$ associated with a compact direction. The 
precise way in which one should deal with such fields has been described in 
\cite{WT-compact}. Briefly stated, transverse fields associated with a 
compact direction can be Fourier decomposed so that they are $T$--dual 
to the momentum modes of the dual gauge field that lives on the 
$T$--dual brane. The Matrix theory expressions for the moments of the 
11--dimensional supergravity currents that we have used in the 
previous section can all be written easily in the 10--dimensional 
language, and this has been done in \cite{Mark-Wati-5}. We refer to 
the appendix for those expressions.

Given the $D0$--brane action in weak background fields, we can now 
write down the $T$--dual action for the Type IIB $D$--string. This 
has in fact been done for any $Dp$--brane \cite{Mark-Wati-5}, as 
briefly discussed in section 2. The $D$--string action is therefore,

\bea \label{sd1nsns}
S^{D1}_{NS-NS} &=& (\phi - {1\over2} h_{\hat{a}\hat{a}}) I_{\phi} + 
{1\over2} h_{00} I^{00}_{h} + {1\over2} h_{ij} I^{ij}_{h} - {1\over2} 
h_{\hat{a}\hat{b}} I^{\hat{a}\hat{b}}_{h} + h_{0i} I^{0i}_{h} + 2 
h_{\hat{a}i} I^{\hat{a}i}_{s} - 2 h_{0\hat{a}} I^{0\hat{a}}_{s} 
\nonumber \\
&
& + 
B_{ij} I^{ij}_{s} - B_{\hat{a}\hat{b}} I^{\hat{a}\hat{b}}_{s} + 2 B_{0i} 
I^{0i}_{s} + B_{\hat{a}i} I^{\hat{a}i}_{h} - B_{0\hat{a}} I^{0\hat{a}}_{h} 
\nonumber \\
&
& + 
{\rm Higher \; moment \; terms} + {\rm Nonlinear \; terms},
\eea

\bea \label{sd1rr}
S^{D1}_{R-R} &=& 
\int d^{2}\sigma \; \epsilon^{a_{0}a_{1}} \sum_{q} 
\sum_{n={\rm Max}(0,q-2)}^{n={\rm Min}(q,8)} {(-1)^{n(n-1)\over 2} 
(2n-q+1)!! \over n! (q-n)! (n-q+2)!} \; \str \; \{ C^{(q)}_{a_{0} \cdots 
a_{q-n-1} i_{1} \cdots i_{n}} \cdot
\nonumber \\
&
&
\cdot F^{n+{2-q \over 2}}_{(a_{q-n} \cdots a_{1} i_{1} \cdots i_{n})} \}
+ {\rm Higher \; moment \; terms} + {\rm Nonlinear \; terms},
\eea

\noindent
where the indices in curved brackets are to be assigned pairwise to 
the corresponding product of $F$'s and then symmetrized over all 
orderings. Indices $0,\hat{a},\ldots$, live on the 2--dimensional 
world--volume, while indices $i,j,\ldots$, are transverse to the 
$D$--string. The $I$ currents in these expressions should not be confused 
with the $I$ currents derived in the previous section. In here we started 
with the $D0$--brane currents mentioned in section 2, which in turn can be 
determined in terms of the Matrix currents $T$, $J$ and $M$. As we 
wrote the previous expressions, the 0--brane currents $I$ are then to 
be reinterpreted as 10--dimensional expressions reduced to the 
2--dimensional world--volume of the $D1$--brane. As to the higher 
moment terms, the expressions will be just like the ones above, but 
with the appropriate inclusion of arbitrary derivatives of each 
background field.

It is therefore clear that in order to explicitly write the 
$D$--string action, it would be useful to start with the 
10--dimensional expressions for the $I$ currents. These can be 
obtained from the expressions in \cite{Mark-Wati-5}, and they are as 
follows. Observe that we write down the expressions in the previously 
explained 10--dimensional notation, and so when reducing to the 
2--dimensional world--volume one should take into consideration the 
compact direction carefully, as was mentioned before.

The $I^{00}_{h}$ component of the matrix string current for 
background metric field, can be written in 10--dimensional notation 
as (we dropped a factor of $1/R$ from all the expressions that follow 
for the $I$ currents):


\bea
I^{00}_{h} &=& T^{++} + T^{+-} + (I^{00}_{h})_{8} + {\cal O}(X^{12}) 
\nonumber \\
&=& \str ( \identity + F^{0\mu}F^{0}{}_{\mu} + {1\over 4} F_{\mu\nu} 
F^{\mu\nu} + {\cal O}(\Theta^{2}) ) 
+ (I^{00}_{h})_{8} + {\cal O}(X^{12}).
\eea

\noindent
All components of these matrix string currents are 
straightforward to write down, so we simply present them. For the 
NS--NS sector, the matrix string current components can be written in 
10--dimensional notation as:


\bea
I^{00}_{h} &=& T^{++} + T^{+-} + (I^{00}_{h})_{8} + {\cal O}(X^{12}) 
\nonumber \\
&=& \str ( \identity + F^{0\mu}F^{0}{}_{\mu} + {1\over 4} F_{\mu\nu} 
F^{\mu\nu} + {\cal O}(\Theta^{2}) ) 
+ (I^{00}_{h})_{8} + {\cal O}(X^{12}),
\nonumber \\
I^{0i}_{h} &=& T^{+i} + T^{-i} + {\cal O}(X^{10})
\nonumber \\
&=& - \str ( F^{0i} + F^{0\mu} F_{\mu\nu} F^{\nu i} + {1\over 4} 
F^{0i} F_{\mu\nu} F^{\mu\nu} + {\cal O}(\Theta^{2},\Theta^{4}) )
+ {\cal O}(X^{10}),
\nonumber \\
I^{ij}_{h} &=& T^{ij} + (I^{ij}_{h})_{8} + {\cal O}(X^{12}) 
= \str(F^{i\mu} F_{\mu}{}^{j} + {\cal O}(\Theta^{2})) 
+ (I^{ij}_{h})_{8} + {\cal O}(X^{12}),
\nonumber \\
I_{\phi} &=& T^{++} - {1\over 3} ( T^{+-} + T^{ii} ) + 
(I_{\phi})_{8} + {\cal O}(X^{12}) 
\nonumber \\
&=& \str( \identity + {1\over 4} F_{\mu\nu} F^{\mu\nu} + 
{\cal O}(\Theta^{2})) + (I_{\phi})_{8} + {\cal O}(X^{12}),
\nonumber \\
I^{0i}_{s} &=& 3 J^{+-i} + {\cal O}(X^{8})
= {1\over 2} \str(F^{0\mu} F_{\mu}{}^{i} + {\cal O}(\Theta^{2}))
+ {\cal O}(X^{8}),
\nonumber \\
I^{ij}_{s} &=& 3 J^{+ij} - 3 J^{-ij} + {\cal O}(X^{10})
\nonumber \\
&=& - {1\over 2} \str(F^{ij} + F^{i\mu} F_{\mu\nu} F^{\nu j} + 
{1\over 4} F^{ij} F_{\mu\nu} F^{\mu\nu} + {\cal 
O}(\Theta^{2},\Theta^{4})) + {\cal O}(X^{10}).
\eea

Moving to the R--R fields, one can write the matrix string currents in 
the 10--dimensional notation as:


\bea
I^{0}_{0} &=& T^{++} = \str(\identity),
\nonumber \\
I^{i}_{0} &=& T^{+i} = - \str(F^{0i}),
\nonumber \\
I^{0ij}_{2} &=& J^{+ij} + {\cal O}(X^{10}) = -{1\over 6} \str(F^{ij}) 
+ {\cal O}(X^{10}),
\nonumber \\
I^{ijk}_{2} &=& J^{ijk} + {\cal O} (X^{8}) 
= {1\over 6} \str( F^{0i} F^{jk} + F^{0j} F^{ki} + F^{0k} 
F^{ij} + {\cal O}(\Theta^{2})) + {\cal O} (X^{8}),
\nonumber \\
I^{0ijkl}_{4} &=& 6 M^{+-ijkl} + {\cal O}(X^{8}) 
= {1\over 2} \str(F^{ij} F^{kl} + F^{ik} F^{lj} + F^{il} 
F^{jk} + {\cal O}(\Theta^{2}))
+ {\cal O}(X^{8}),
\nonumber \\
I^{ijklm}_{4} &=& - 6 M^{-ijklm} + {\cal O}(X^{10}) 
= {15\over 2} \str(F^{0[i} F^{jk} F^{lm]} + {\cal O}(\Theta^{2}))
+ {\cal O}(X^{10}),
\nonumber \\
I^{0ijklmn}_{6} &=& S^{+ijklmn} + {\cal O}(X^{10}) 
= \str\left(F_{[ij} F_{kl} F_{mn]}\right) + {\cal O}(X^{10}),
\nonumber \\
I^{ijklmnp}_{6} &=& S^{ijklmnp} + {\cal O}(X^{12}) 
= 7 \, \str\left(F_{[ij} F_{kl} F_{mn} \dot{X}_{p]} + 
{\cal O}(\theta^{2},\theta^{4}) \right) + {\cal O}(X^{12}).
\eea

\noindent
As we discussed in the previous section, finding a Matrix theory form 
for the transverse $M5$--brane current components $M^{+ijklm}$ and 
$M^{ijklmn}$ is a matter which is yet not quite fully understood.

All these straighten out, we can proceed and explicitly write down 
(\ref{sd1nsns}) and (\ref{sd1rr}) for this case of the IIB 
$D$--string. From the previous expressions, one obtains:

\bea\label{sdsns}
S^{D1}_{NS-NS} &=& (\phi - {1\over2} h_{a}{}^{a}) \, 
\str \left( \identity + {1\over 4} F_{\mu\nu} F^{\mu\nu} 
+ \cdots \right)
\nonumber \\
&
& + {1\over2} h_{ij} \, 
\str \left( F^{i\mu} F_{\mu}{}^{j} + \cdots \right) - {1\over2} 
h_{ab} \, \str \left (F^{a\mu} F_{\mu}{}^{b} + \cdots \right)
\nonumber \\
&
& - h_{ai} \,
\str \left( F^{ai} + F^{a\mu} F_{\mu\nu} F^{\nu i} + {1\over 4} F^{ai} 
F_{\mu\nu} F^{\mu\nu} + \cdots \right)
\nonumber \\
&
& - {1\over 2} B_{ij} \,
\str \left( F^{ij} + F^{i\mu} F_{\mu\nu} F^{\nu j} + 
{1\over 4} F^{ij} F_{\mu\nu} F^{\mu\nu}  + \cdots \right)
\nonumber \\
&
& + {1\over 2} B_{ab} \,
\str \left( F^{ab} + F^{a\mu} F_{\mu\nu} F^{\nu b} + 
{1\over 4} F^{ab} F_{\mu\nu} F^{\mu\nu} + \cdots \right)
\nonumber \\
&
& + B_{ai} \,
\str \left( F^{a\mu} F_{\mu}{}^{i} + \cdots \right) + \ldots,
\eea

\bea \label{sdsrr}
S^{D1}_{R-R} &=& {1\over 2} \epsilon^{ab} C^{D(-1)} \, \str \left( 
F_{ab} + \cdots \right) 
+ {1\over 2} \epsilon^{ab} C^{D1}_{ab} \, \str \left( \identity + 
\cdots \right) + {1\over 2} \epsilon^{ab} C^{D1}_{ai} \, \str 
\left( F_{b}{}^{i} + \cdots \right) 
\nonumber\\
&
& + {1\over 2} \epsilon^{ab} C^{D1}_{ij} \, \str \left( F_{a}{}^{i} 
F_{b}{}^{j} - {1\over 2} F_{ab} F^{ij} + \cdots \right) + \ldots,
\eea

\noindent
where in the previous expressions one should still take into 
consideration that the tensors must be appropriately reduced to 
2--dimensional world--volume notation via the identifications $F_{ai} \equiv 
D_{a} X^{i}$ and $F_{ij} \equiv i[X^{i},X^{j}]$, and the proper 
treatment of the compact coordinate $X^{9}$ according to 
\cite{WT-compact}. Integration over the cylindrical world--sheet 
$\{\tau,\sigma\}$ is implicit. Of course the action (\ref{sdsns}) 
and (\ref{sdsrr}) is quite interesting on its own, as it yields the 
gauged matrix sigma model for the Type IIB $D$--string in weakly 
curved backgrounds.

The issue of $T$--duality along $R_{9}^{IIA}$ solved, let us now deal 
with the IIB $S$--duality transformation. The $SL(2,{\bf Z})$ duality 
symmetry of Type IIB string theory maps a $(p,q)$--string into another 
$(p',q')$--string. In here, we shall focus on the usual ${\bf Z}_{2}$ 
subgroup of the $S$--duality group generated by the transformation 
which exchanges the NS--NS and R--R 2--form fields, and in 
particular maps the $D$--string into the fundamental string. As in 
the case of $T$--duality, the action of this subgroup of $S$--duality 
on arbitrary IIB supergravity background fields is well known 
\cite{Bergshoeff-Hull-Ortin}. At linear order the transformation is 
\cite{Mark-Wati-5},

\bea \label{lsdt}
\phi &\longrightarrow& - \phi
\nonumber \\
C^{(0)} &\longrightarrow& - C^{(0)}
\nonumber \\
B_{\mu\nu} &\longrightarrow& - C^{(2)}_{\mu\nu}
\nonumber \\
C^{(2)}_{\mu\nu} &\longrightarrow& B_{\mu\nu}
\nonumber \\
h_{\mu\nu} &\longrightarrow& h_{\mu\nu}
\nonumber \\
C^{(4)} &\longrightarrow& C^{(4)}.
\eea

\noindent
These transformations are written in the Einstein frame, even though 
we have been working in the string frame. This is fine as the terms that 
we are considering from the string action, in this paper, are the same in 
both frames.

The problem we face concerns the implementation of $S$--duality at 
the $D$--string world--volume level. In fact, the $S$--duality 
transformation properties of the world--volume operators in the 
2--dimensional $U(N)$ gauge theory are not known. What we shall 
see, is that in the end the $S$--duality transformations are not so 
surprising, and they turn out to be quite simple as there will be no 
change in the composite operators. But for the moment, let us assume 
that they could be anything.

In here, for the $D1$--brane, because we know what the result is from 
matrix string theory we can predict the precise transformation properties 
of all the operators that appear in the action. For the moment, we 
shall perform the IIB $S$--duality transformation of the $D$--string 
action (\ref{sdsns}) and (\ref{sdsrr}) in the following way. We apply 
the linear $S$--duality transformations (\ref{lsdt}) to the background 
fields, and we denote $S$--duals of the world--volume fields by simply 
putting a tilde over them. Next, as we $T$--dualize back to the IIA 
theory, we can then compare to the previous section and obtain 
predictions for all these ``tilded'' operators. $S$--dualizing the 
$D$--string action in this way, one obtains:

\bea\label{sfsns}
S^{F1}_{\widetilde{NS-NS}} &=& (-\phi - {1\over2} h_{a}{}^{a}) \, 
\str \left( \identity + {1\over 4} \widetilde{F_{\mu\nu} F^{\mu\nu}} 
+ \cdots \right)
\nonumber \\
&
& + {1\over2} h_{ij} \, 
\str \left( \widetilde{F^{i\mu} F_{\mu}{}^{j}} + \cdots \right) - {1\over2} 
h_{ab} \, \str \left( \widetilde{F^{a\mu} F_{\mu}{}^{b}} + \cdots \right)
\nonumber \\
&
& - h_{ai} \,
\str \left( \widetilde{F^{ai}} + \widetilde{F^{a\mu} F_{\mu\nu} F^{\nu i}} + 
{1\over 4} \widetilde{F^{ai} F_{\mu\nu} F^{\mu\nu}} + \cdots \right)
\nonumber \\
&
& + {1\over 2} C_{ij} \,
\str \left( \widetilde{F^{ij}} + \widetilde{F^{i\mu} F_{\mu\nu} F^{\nu j}} + 
{1\over 4} \widetilde{F^{ij} F_{\mu\nu} F^{\mu\nu}}  + \cdots \right)
\nonumber \\
&
& - {1\over 2} C_{ab} \,
\str \left( \widetilde{F^{ab}} + \widetilde{F^{a\mu} F_{\mu\nu} F^{\nu b}} + 
{1\over 4} \widetilde{F^{ab} F_{\mu\nu} F^{\mu\nu}} + \cdots \right)
\nonumber \\
&
& - C_{ai} \,
\str \left( \widetilde{F^{a\mu} F_{\mu}{}^{i}} + \cdots \right)
+ \ldots,
\eea

\bea \label{sfsrr}
S^{F1}_{\widetilde{R-R}} &=& - {1\over 2} \epsilon^{ab} C^{D(-1)} \, \str 
\left( \widetilde{F_{ab}} + \cdots \right) 
+ {1\over 2} \epsilon^{ab} B_{ab} \, \str \left( \identity + 
\cdots \right) + {1\over 2} \epsilon^{ab} B_{ai} \, \str \left( 
\widetilde{F_{b}{}^{i}} + \cdots \right) 
\nonumber\\
&
& + {1\over 2} \epsilon^{ab} B_{ij} \, \str \left( \widetilde{F_{a}{}^{i} 
F_{b}{}^{j}} - {1\over 2} \widetilde{F_{ab} F^{ij}} + \cdots \right) + \ldots,
\eea

\noindent
where we also denoted the action subscripts with a tilde, as we have a 
mixing of the NS--NS and R--R sectors under $S$--duality. A note on 
notation: in case it is not clear, in the previous expressions the tilde 
is over the whole composite operator.

Proceeding, we are left with a final $T$--duality along $R_{9}^{IIB}$ 
that leads back to the Type IIA theory, and therefore to matrix 
string theory. The rules for $T$--duality have been previously 
explained and used, so we just apply them to the previous expressions 
to obtain the matrix string theory action in weak background fields. 
Two points should still be stressed. We have been slightly abusive of 
notation in the previous expressions by allowing more than one 
compact direction, as we indexed tensors as $\hat{a},\hat{b},\ldots$. 
Of course in the case we are dealing with there is only one, 
$\sigma$. Moreover, we still have to write the world--volume tensors in 
the 2--dimensional world--sheet notation. Once we make the notation 
completely rigorous, we are left with (recall that in the matrix 
string limit $R_{9} \to 0$),

\bea \label{mstatst}
{\cal S} &=& \phi \, 
\str \left( - \identity + {1\over2} \widetilde{\dot{X}_{i}^{2}} - {1\over2} 
\widetilde{(DX^{i})^{2}} + {1\over2} g_{s}^{2} \widetilde{\dot{A}^{2}} + 
{1\over 2g_{s}^{2}} \sum_{i<j} \widetilde{[X^{i}, X^{j}]^{2}} + \cdots \right) 
\nonumber \\
&
& + {1\over 2} h_{00} \, 
\str \left( \identity + {1\over2} \widetilde{\dot{X}_{i}^{2}} + {1\over2} 
\widetilde{(DX^{i})^{2}} + {1\over2} g_{s}^{2} \widetilde{\dot{A}^{2}} - 
{1\over 2g_{s}^{2}} \sum_{i<j} \widetilde{[X^{i}, X^{j}]^{2}}
+ \cdots \right)
\nonumber \\
&
& + {1\over2} h_{ij} \, 
\str \left( \widetilde{\dot{X}^{i}\dot{X}^{j}} - \widetilde{(DX^{i})(DX^{j})} 
- {1\over g_{s}^{2}} \widetilde{[X^{i}, X^{k}][X^{k}, X^{j}]} 
+ \cdots \right) 
\nonumber \\
&
& + h_{0i} \,
\str ( \widetilde{\dot{X}^{i}}
+ {1\over2} g_{s}^{2} \widetilde{\dot{A}^{2} \dot{X}^{i}} - i 
\widetilde{\dot{A} (DX^{j}) [X^{j}, X^{i}]} + {1\over2} \widetilde{
(\dot{X}^{j})^{2} \dot{X}^{i}} - \widetilde{\dot{X}^{j} (DX^{j}) 
(DX^{i})} 
\nonumber \\
&
& + {1\over2} 
\widetilde{(DX^{j})^{2} \dot{X}^{i}} - {1\over g_{s}^{2}} \widetilde{ 
\dot{X}^{j} [X^{j}, X^{k}] [X^{k}, X^{i}]}  - {1\over 2g_{s}^{2}} 
\widetilde{\dot{X}^{i} \sum_{j<k} [X^{j}, X^{k}]^{2}} + \cdots ) 
\nonumber \\
&
& + B_{0i} \, 
\str \left( - {1\over 2} \widetilde{(DX^{i})} + \cdots \right) 
+ B_{ij} \, \str ( - {1\over 2} \widetilde{\dot{X}^{i}(DX^{j})} 
+ {1\over 2} \widetilde{\dot{X}^{j}(DX^{i})} - {i\over 2} 
\widetilde{\dot{A}[X^{i}, X^{j}]} 
\nonumber \\
&
& + \cdots ) + 
C_{i} \, \str \left( g_{s} \widetilde{\dot{A}\dot{X}^{i}} - {i\over 
g_{s}} \widetilde{(DX^{j})[X^{j}, X^{i}]} + \cdots \right) 
+ C_{0} \, \str ( g_{s} \widetilde{\dot{A}} + {1\over 2} g_{s}^{3} 
\widetilde{\dot{A}^{3}} 
\nonumber \\
&
& - {1\over 2} g_{s} 
\widetilde{\dot{A} (DX^{i})^{2}} + {1\over 2} g_{s} \widetilde{ 
\dot{A} (\dot{X}^{i})^{2}} + {i\over g_{s}} \widetilde{ 
\dot{X}^{i} [X^{i}, X^{j}] DX^{j}} - {1\over 2g_{s}} \widetilde{
\dot{A} \sum_{i<j} [X^{i}, X^{j}]^{2}} + \cdots ) 
\nonumber \\
&
& + \ldots
\eea

\noindent
This completes the sequence of DVV dualities. The final result should 
be equivalent to performing the $9-11$ flip of section 4. Comparing 
this result to the one in section 4, (\ref{mstawcb}), one will obtain 
the $S$--duality transformation rules for the tensor operators 
that live on the $D$--string world--volume. One should only keep in 
mind that in this section we have dropped a factor of $1/R$ from all 
the expressions, and that one may need to correct for units as 
comparing to the previous section.

From the expression (\ref{mstatst}) we have results for the 
$I_{\phi}$, $I^{00}_{h}$, $I^{ij}_{h}$, $I^{0i}_{h}$, $I^{0i}_{s}$, 
$I^{ij}_{s}$, $I^{i}_{0}$ and $I^{0}_{0}$ tensors coming from the 
$T$--$S$--$T$ chain of dualities. It should be a straightforward 
exercise to include all other matrix string tensors in this result. 
For our purposes this is enough. Comparing back to what we have 
obtained for those same tensors in section 4 -- where we used the 
$9-11$ flip to DVV reduce -- we obtain an interesting result: the 
2--dimensional $D$--string world--volume composite operators are invariant 
under the target space IIB $S$--duality operation. It is indeed somewhat 
expected that these operators should not change, as from the field 
theory side we expect non--trivial $S$--duality properties only for 
the ${\cal N} = 4$, $d=4$ gauge theory, {\it i.e.}, we only expect to see 
non--trivial transformation laws for the operators that live on the 
world--volume of the $D3$--brane. For the $D1$--brane, the operators 
are kept fixed under the target transformation.

\section{Green--Schwarz Action in a Curved Background}

Given that the matrix string theory action we have built has been 
firmly established at the level of string duality, we would further 
like to confirm it by looking at its conformal field theory limit, 
where we obtain the free string case. We expect that when we take the 
$g_{s} \to 0$ limit, our action should match the Green--Schwarz sigma 
model for the Type IIA superstring \cite{Fradkin-Tseytlin}, once we 
consider this latter one in the same weak background approximation 
that we are considering in here.

\subsection{The Green--Schwarz Action}

We begin with a short review of the results obtained for the IIA 
superstring in \cite{Fradkin-Tseytlin}, so that we can establish a 
bridge between the results of that paper and our notation. We want to 
compare the Green--Schwarz sigma model to our matrix sigma model, and for 
that all one needs to do is to consider the matrix sigma model in the free 
string limit which was presented throughout section 4 for all the tensor 
fields. One should also take into consideration the weak field approximation 
in the sigma model for one string. It should be stressed that we shall be 
looking at schematic and qualitative results only, throughout this section. 
This is because establishing a precise map from the matrix string theory 
in the IR to the light--cone Green--Schwarz action requires a precise 
lifting of the IR matrix action from the cylinder to its branched coverings, 
and so to any given Riemann surface. The procedure is described at length 
in \cite{Bonelli-Bonora-Nesti-1, Bonelli-Bonora-Nesti-2, 
Bonelli-Bonora-Nesti-Tomasiello-1, Bonelli-Bonora-Nesti-Tomasiello-2, 
Bonelli} for the flat background situation. Completing such a 
procedure for this curved situation is an interesting project for the 
future.

The massless spectra of the Type IIA closed superstring includes the 
metric, $g_{\mu\nu}$, the NS $B$--field, $B_{\mu\nu}$, and the 
dilaton, $\phi$, whose origin is the bosonic sector of the 
superstring action. It also includes the $D0$--brane 1--form, 
$C_{\mu}$, and the $D2$--brane 3--form, $C_{\mu\nu\rho}$, whose origin 
is the fermionic sector of the same superstring action. The covariant 
superstring action can be written while in the presence of couplings to the 
background fields of ${\cal N}=2$ 10--dimensional supergravity, as 
was shown in \cite{Fradkin-Tseytlin}. In here we are interested in 
the form of this action in light--cone gauge, which was also derived in 
\cite{Fradkin-Tseytlin}: If one chooses light--cone gauge, and 
furthermore assumes the supergravity background fields to be 
non--trivial only in the eight transverse directions (so that the 
background spacetimes decomposes as ${\cal M}^{10} = {\bf R}^{(1,1)} 
\times {\cal M}^{8}$), then the NS--NS sector of the IIA superstring 
action is written as \cite{Fradkin-Tseytlin},

\bea \label{ssnsns}
{\cal L}_{NS-NS} &=& g_{ij}(X) \; \sigma^{ab} \partial_{a} X^{i} \partial_{b} 
X^{j} + 4 \pi \alpha' \; B_{ij} (X) \; \epsilon^{ab} \partial_{a} 
X^{i} \partial_{b} X^{j} - 2i \theta^{T} \gamma^{0} \rho^{\hat{0}} 
\gamma^{-} \rho^{a} \widehat{D}_{a} \theta
\nonumber \\
&
& + {1\over 64} \widehat{R}_{ijkl} \; \theta^{T} \gamma^{0} 
\rho^{\hat{0}} \gamma^{ij-} \rho^{a} (1+\rho_{3}) \theta \; 
\theta^{T} \gamma^{0} \rho^{\hat{0}} \gamma^{kl-} \rho^{a} 
(1-\rho_{3}) \theta + \cdots,
\eea

\noindent
where we have the following relations,

\bea
\widehat{R}^{i}{}_{jkl} &=& \partial_{k} \widehat{\Gamma}^{i}_{jl} - \cdots, 
\;\;\;\;\;\;  \widehat{\Gamma}^{i}_{jl} = \Gamma^{i}_{jl} [g] + 2\pi 
\alpha' H^{i}{}_{jl}, \;\;\;\;\;\; H_{ijk} = 3\partial_{[i} B_{jk]},
\nonumber \\
\widehat{D}_{a} &=& \partial_{a} - {1\over4} \gamma^{\hat{j}\hat{k}} 
\, \widehat{\omega}_{i}^{\hat{j}\hat{k}} \, \partial_{a} X^{i}, 
\;\;\;\;\;\; \widehat{\omega}_{..i} = \omega_{..i} + 2\pi\alpha' 
\rho_{3} H_{..i}, \;\;\;\;\;\; \gamma^{ij-} \equiv \gamma^{ij} \gamma^{-},
\eea

\noindent
and moreover: $i,j,k,\ldots$, are transverse spacetime indices; 
$a,b,\ldots$ are world--sheet indices; hatted indices correspond to 
tangent frame indices; $\sigma_{ab}$ is the world--sheet metric; and we 
have introduced two--dimensional world--sheet Dirac matrices 
$\rho_{\hat{a}}$ as,

\be
\rho^{\hat{0}} = \left(
\begin{array}{cc}
0 & -i \\
i & 0 
\end{array} \right), \;\;\;\;\;\;
\rho^{\hat{1}} = \left(
\begin{array}{cc}
0 & i \\
i & 0 
\end{array} \right), \;\;\;\;\;\;
\rho_{3} = \left(
\begin{array}{cc}
1 & 0 \\
0 & -1 
\end{array} \right).
\ee

\noindent
One can also write down the Lagrangian for the R--R sector of the Type 
IIA superstring action \cite{Fradkin-Tseytlin},

\bea \label{ssrr}
{\cal L}_{R-R} &=& {i\over (\alpha')^{3/2}} \; \theta^{T} \gamma^{0} 
\rho^{\hat{0}} \gamma^{(i} \Gamma^{\Lambda} \gamma^{j)} (1-\rho_{3}) 
\theta \; ( - \partial^{a} X^{i} \partial_{a} X^{j} + {i\over6} 
\theta^{T} \gamma^{0} \rho^{\hat{0}} \gamma^{ni-} \rho^{a} \theta 
\; \partial_{a} X^{j} \partial_{n} 
\nonumber \\
&
& - {1\over144} \;
\theta^{T} \gamma^{0} \rho^{\hat{0}} \gamma^{ni-} \rho^{a} \theta \; 
\theta^{T} \gamma^{0} \rho^{\hat{0}} \gamma^{mj-} \rho_{a} \theta 
\; \partial_{n} \partial_{m} ) \; C_{\Lambda} (X) 
\nonumber \\
&
& + {i\over (\alpha')^{3/2}} \; \theta^{T} \gamma^{0} 
\rho^{\hat{0}} \gamma^{[i} \Gamma^{\Lambda} \gamma^{j]} (1-\rho_{3}) 
\theta \; ( - \epsilon^{ab} \partial_{a} X^{i} \partial_{b} X^{j} + 
{i\over6} \theta^{T} \gamma^{0} \rho^{\hat{0}} \gamma^{ni-} \rho^{a} 
\rho_{3} \theta \; \partial_{a} X^{j} \partial_{n} 
\nonumber \\
&
& - {1\over144} \;
\theta^{T} \gamma^{0} \rho^{\hat{0}} \gamma^{ni-} \rho^{a} \theta \; 
\theta^{T} \gamma^{0} \rho^{\hat{0}} \gamma^{mj-} \rho_{a} \rho_{3} 
\theta \; \partial_{n} \partial_{m} ) \; C_{\Lambda} (X) + \cdots,
\eea

\noindent
where we have $\{ ( \Gamma^{\Lambda}, C_{\Lambda} ) \} = \{ ( \gamma^{-i}, 
C_{i} ), ( \gamma^{ijk-}, C_{ijk} ) \}$ for the background $D0$ and 
$D2$--brane currents. Indices $i,j,\ldots$, are contracted via the 
metric $g_{ij}$, and the dots in (\ref{ssrr}) refer to 
higher--derivative terms (which have no contribution to the 
light--cone vertex operators) \cite{Fradkin-Tseytlin}. This completes 
the information on the Green--Schwarz action that we shall be 
interested in.

To begin the comparison with the abelian limit of our results from 
section 4, we look at the NS--NS sector. In the weak background field 
limit we are considering, the Riemann curvature terms drops out from 
(\ref{ssnsns}) and all we need to check is for the existence of the 
abelian couplings,

\be
h_{ij}(X) \; \partial^{a} X^{i} \partial_{a} X^{j},
\ee

\noindent
and

\be
B_{ij} (X) \; \epsilon^{ab} \partial_{a} X^{i} \partial_{b} X^{j}.
\ee

\noindent
Comparing back to (\ref{cftih}) and (\ref{cftis}), one immediately 
observes that these terms indeed appear in the abelian conformal limit 
of our matrix string action. The term involving the Riemann tensor is 
present in a weak field approximation only via its piece in $\partial^{2} h$. 
One can realize however that, being schematically of the form 
$\widehat{R}_{ijkl} \;\theta^{T} \Gamma^{ij} \theta \; \theta^{T} \Gamma^{kl}
\theta$, it is of order ${\cal O} = ({R\over \ell_{s}})^{2}$. At the Matrix 
theory level we have four fermion terms of this type in the $T^{-i}$ 
component of the Matrix stress tensor and in the $J^{-ij}$ component 
of the Matrix membrane current, and therefore we have terms of this 
type in several components of the matrix string tensor couplings. 
However, none of these terms seems to have the required index structure to 
couple to the Riemann curvature term, and besides we are looking for 
tensors that will couple to a term of the type $\partial_{i}\partial_{j} 
h_{kl}$. The reason for this is that such couplings will actually arise from 
higher moment terms. Indeed, one can see from the appendix that there 
are two fermion contributions to the first moment terms of 
$T^{ij(l)}_{\rm Fermion}$, which will couple to a term in $\partial 
h$. Similarly there will be four fermion contributions that should 
produce the required curvature coupling. It would be interesting to 
construct explicitly such terms.

When we move to the R--R sector, we observe that in (\ref{ssrr}) the 
terms that do not involve derivatives of the $D0$ and $D2$--brane 
fields are at order ${\cal O} = ({R\over \ell_{s}})^{2}$, and 
schematically of the form $\theta^{T} \Gamma_{ij\Lambda} \theta \; \partial
X^{i} \partial X^{j} $. For the $D0$--brane such terms will likely 
come from the quadratic fermion pieces that we neglected in the 
tensor $T^{--}$, while for the $D2$--brane case it is not entirely 
clear where these terms should come form. The other terms in the R--R action 
(\ref{ssrr}) are higher derivative terms in the R--R fields, and so we would 
only expect to match these terms to higher moments of our couplings. Therefore, 
the overall comparison of our results with the ones for the one string 
action is rather schematic and qualitative. But the comparison can 
still be of some use in predicting some possible new coupling terms 
coming up in the full curved action.

\subsection{Matrix Theory in Curved Backgrounds}

As we mentioned before, in the Green--Schwarz sigma model there are 
four fermion couplings to the background Riemann tensor. One 
could expect that this coupling would correspond to the free string 
limit of some nonabelian coupling between world--sheet fields and the 
background curvature. Indeed, given the previous match between abelian 
and nonabelian actions, one has some clues for the form of the couplings 
to background curvature. This would amount to terms in the full 
non--linear matrix string action, and therefore to terms involving 
the Riemann curvature and other non--linear combinations of the 
supergravity background fields in the Matrix theory action.

As we have just seen, the term involving the curvature tensor is 
schematically of the form $\widehat{R}_{ijkl} \;\theta^{T} \Gamma^{ij} 
\theta \; \theta^{T} \Gamma^{kl} \theta$, and we have no tensor 
coupling with this index structure among the zeroeth moment terms. One 
naturally expects that such tensors would start making their appearance once 
one performs a higher loop calculation in Matrix theory or in matrix string 
theory as we would obtain, {\it e.g.}, quadratic pieces in the 
metric, $h$. For the moment we simply observe that the actions (\ref{ssnsns}) 
and (\ref{ssrr}) yield already some information on what one will obtain from 
such a higher loop calculation by telling us what will be the abelian limit 
of the tensors one would eventually obtain. Indeed we can predict 
that there will be a tensor coupling to the quadratic metric piece, of the type,

\be
G^{ijkl} = \theta^{T} \Gamma^{ij} \theta \; \theta^{T} \Gamma^{kl} 
\theta + {\rm Nonabelian} \; {\rm Terms}.
\ee

\noindent
This is of course the required coupling for the curved matrix string 
theory action to match, in the free abelian limit, the Green--Schwarz 
action. But because we are dealing in this section with background 
fields that are non--trivial only in the eight transverse directions, 
this term actually is lifted to a term in the curved Matrix theory 
action. We see therefore that in order for a supersymmetric completion 
of the curved background Matrix action to be done, there will be at 
least quartic fermion terms required, at zeroeth moment terms.

Because the curvature term in (\ref{ssnsns}) also includes a coupling 
to the NS--NS $B$--field field strength, a similar story will also 
work for that term. Likewise, due to the higher derivative terms in 
(\ref{ssrr}), we shall actually obtain quite a few predictions for 
higher moments and other tensor couplings, along the lines of the 
previous discussion for the background Riemann curvature. In summary, 
a full extension of Matrix theory for the case of curved backgrounds 
will probably not be as simple as the sigma model proposal in 
\cite{Douglas-talk}. Its supersymmetric completion however, will 
have to include a series of non--linear background couplings, as we 
have just discussed. Such types of couplings, involving the Riemann tensor
and four fermion fields, are common in supersymmetric completions of
bosonic sigma models and so are quite natural to be expected in here as
well.

\section{Noncommutative Backgrounds}

In this section we wish to exemplify the nonabelian nature of our 
action. Ideally one hopes that coherent states of gravitons can be 
made out of many fundamental strings (by some sort of fundamental string 
condensation), and by using infinite dimensional matrices to describe such 
solutions one could then be able to build fully curved spacetime geometries 
-- somewhat like when one uses infinite matrices to describe non--compact 
curved membranes in Matrix theory \cite{Lorenzo-Wati}. In practice the 
situation is not as idealistic, as it is not clear how to go from the 
description of the coherent state in terms of the strings to their effects 
on the other strings. This would correspond to a higher loop 
calculation in Matrix theory or matrix string theory.

Still, an interesting question is whether one can exponentiate the 
noncommutative vertex operators we have obtained in our linear action, 
and from that build the full non--linear matrix sigma model. Recall that 
the results for the $I$ tensors can be used (loosely speaking, via 
multiplication by $\exp (ip \cdot X)$) to obtain the noncommutative vertex 
operators of matrix string theory. However, precisely because of this 
noncommutative nature of the vertex operators, one still lacks an 
ordering for the exponentiation. Such an ordering would moreover have 
to produce terms with derivatives of the background fields, as such 
terms are expected in order to satisfy the geodesic length condition 
in \cite{Douglas-curved,dko}. It is certainly not clear how to choose the 
ordering of such an exponential at this stage. There is also the question 
of whether the background satisfies the equations of motion of 
supergravity. To clear this issue, one would again need the full matrix 
sigma model in order to compute noncommutative beta functions and from 
there derive the noncommutative equations of motion for the background.

For the moment we will aim lower and consider a very simple example involving 
non--trivial R--R flux. There is a particular interest in examples 
involving R--R flux, due to its possible connection to noncommutative 
spacetime geometry. Recently there has been some study in the applications 
of noncommutative geometry to string theory. This has, however, been 
mainly studied at the world--volume level where the noncommutativity 
appears as a result of non--trivial NS--NS flux 
\cite{Connes-Douglas-Schwarz, Lorenzo-Ricardo, Seiberg-Witten, Lorenzo}. 
But it has also been suggested that in situations involving R--R flux 
rather than NS--NS flux, the noncommutativity could make its appearance 
at the background spacetime level due to small distance stringy effects 
\cite{Chu-Ho-Kao}. It would be quite interesting to further study this 
issue.

\subsection{R--R Flux and String Condensation}

An interesting situation is the one where there exists non--trivial R--R 
flux. In recent work \cite{Myers,Constable-Myers-Tajford} it was studied 
an example where a collection of $D0$--branes was polarized into a 
noncommutative 2--sphere configuration by an external R--R field. The question 
quickly arises of whether a similar situation could exist in this case, where 
we are dealing with a collection of fundamental strings. If they can indeed be 
polarized into noncommutative configurations by some external R--R fields, this 
would then correspond to the creation of some sort of noncommutative 
stringy object. We shall see that such indeed happens, and so R--R flux can 
act as a source for fundamental string world--sheet noncommutativity.

Let us consider a situation where there is non--trivial R--R 3--form 
flux. This case will be quite similar to the one of dielectric branes 
considered in \cite{Myers}, the difference being that now we have 
``dielectric strings''. We will moreover consider a simplified case 
where we take all fermionic fields to vanish, $\theta=0$. As an 
ansatz, let us also consider $A=0$ and $\partial X^{i}=0$. At the 
background level, we shall set all other fields (except for the membrane 
current) to zero. All this done, one is left with the flat space 
matrix string theory action,

\be
S_{Flat} = {1\over 2\pi} \int d\sigma d\tau \; \tr \left( {1\over 2} 
(\dot{X}^{i})^{2} + {1\over 4 g_{s}^{2}} [X^{i}, X^{j}]^{2} \right),
\ee

\noindent
supplemented by the $D2$--brane linear coupling,

\be
S_{D2-brane} = {1\over 2\pi} \int d\sigma d\tau \; 
C_{\mu\nu\lambda} (X) I_{2}^{\mu\nu\lambda}.
\ee

As to the $D2$--brane linear coupling, we shall focus our attention in 
the lowest order terms in the derivative expansion. In particular, we 
will not retain terms at order ${\cal O}=({R\over \ell_{s}})$ and 
above (this corresponds to two operator insertions and is the same 
order as the flat space matrix string action). In this case, the 
tensor couplings we need to consider are:

\bea
I_{2}^{0ij} &=& \str \left( -{i\over 6g_{s}} \left({\ell_{s}\over R}\right) 
[X^{i}, X^{j}] + {\cal O}({R\over\ell_{s}}) \right),
\nonumber \\
I_{2}^{ijk} &=& \str \left( - {i\over 6g_{s}} \dot{X}^{i} [X^{j},X^{k}] 
- {i\over 6g_{s}} \dot{X}^{j} [X^{k},X^{i}] - {i\over 6g_{s}} \dot{X}^{k} 
[X^{i},X^{j}] \right), 
\eea

\noindent
so that the $D2$--brane linear coupling term becomes,

\bea
{1\over 2\pi} \int d\sigma d\tau \; C_{\mu\nu\lambda} (X) 
I_{2}^{\mu\nu\lambda} &=& {1\over 2\pi} \int d\sigma d\tau \; \left( 
3 C_{0ij} (X) I_{2}^{0ij} + C_{ijk} (X) I_{2}^{ijk} 
\right) 
\nonumber \\
&=& {1\over 2\pi} \int d\sigma d\tau \; \str ( 
- {i\over 2g_{s}} \left( {\ell_{s}\over R} \right) C_{0ij} (X) 
[X^{i}, X^{j}] 
\nonumber \\
&
& - {i\over 2g_{s}} C_{ijk} (X) \dot{X}^{i} [X^{j}, 
X^{k}] + {\cal O}({R\over \ell_{s}}) ),
\eea

\noindent
where we still have to expand $C_{0ij}$ to first order in 
derivatives, $C^{D2}_{0ij} (X) = C^{D2}_{0ij} (0) + ({R\over\ell_{s}}) 
X^{k} \partial_{k} C^{D2}_{0ij} (0) + \cdots$. Doing this and 
integrating the time derivative by parts, one obtains,

\be
{1\over 2\pi} \int d\sigma d\tau \; C_{\mu\nu\lambda} (X) 
I_{2}^{\mu\nu\lambda} = {1\over 2\pi} \int d\sigma d\tau \;
{i\over2g_{s}} \; \str \left( (\partial_{0} C_{ijk} - 
\partial_{k} C_{0ij}) X^{k} [X^{i}, X^{j}] \right) + \cdots,
\ee

\noindent
where we have further specialized for the particular case of time 
independent solutions, $\dot{X}^{i}=0$. Any solutions we shall obtain 
will correspond to static backgrounds. Now, the $D2$--brane 3--form 
potential, ${\bf C}_{3}$, is related to the $D2$--brane 4-form field 
strength, ${\bf F}_{4}$, by the standard relation ${\bf F}_{4} = 
d{\bf C}_{3}$, and so one simply has the expected gauge invariant 
coupling:

\be
{1\over 2\pi} \int d\sigma d\tau \; C_{\mu\nu\lambda} (X) 
I_{2}^{\mu\nu\lambda} = {1\over 2\pi} \int d\sigma d\tau \;
{i\over6g_{s}} \; \str \left( F_{0ijk} X^{k} [X^{i}, X^{j}] \right) + 
\cdots.
\ee

\noindent
If we furthermore choose the $D2$--brane field strength as constant, 
$F_{0ijk} = -2F \epsilon_{ijk}$ for $i,j,k=1,2,3$, and zero 
otherwise, we are led into a situation very similar to the $D0$--brane 
case of \cite{Myers}, only now with $F$--strings. Indeed, the 
effective potential we have obtained for the static configuration of 
$N$ fundamental strings is,

\be
V_{eff}(X) = - {1\over 4g_{s}^{2}} \, \tr [X^{i}, X^{j}]^{2} - {i\over 
6g_{s}} \, F_{0ijk} \, \str [X^{i}, X^{j}] X^{k}.
\ee

\noindent
with the following equation for the extrema,

\be
[\, [X^{i}, X^{j}], X^{j}] -{i\over2} g_{s} F_{0ijk} [X^{j}, X^{k}] = 0.
\ee

The case of commuting matrices, $[X^{i},X^{j}]=0$, is a solution with zero 
potential energy. This corresponds to the free string limit, {\it 
i.e.}, it corresponds to a situation describing separated, straight 
and static free strings. More interesting to us would be a noncommuting 
solution. Following \cite{Wati-sphere, Myers}, we consider the ansatz 
$X^{i} = \varphi \sigma^{i}$, $i=1,2,3$, where $\varphi$ is a constant 
and $\sigma^{i}$ are some $N$--dimensional matrix representation of the 
$su(2)$ algebra,

\be \label{su2}
[\sigma^{i}, \sigma^{j}] = 2i\epsilon^{ijk}\sigma^{k}.
\ee

\noindent
Using this ansatz in the equation for the extrema of the string 
effective potential, one immediately obtains that this is indeed a 
solution once we set the constant to be $\varphi = {1\over2} g_{s} F$. If 
we moreover consider the $\sigma^{i}$ as an irreducible representation, 
one computes the Casimir as $\tr (\sigma^{i})^{2} = {1\over3} N (N^{2}-1)$, 
and so the nonabelian solution has an effective potential of,

\be
V_{NC} = - {g_{s}^{2} F^{4} \over 48} N (N^{2} - 1),
\ee

\noindent
{\it i.e.}, the noncommutative solution has an energy lower than the 
commuting one. This means of course that the configuration 
corresponding to separated static free strings is unstable, and the 
strings will actually condense into a noncommutative solution. This solution 
is the noncommutative sphere, with radius ${\sl R} = {1\over2} 
g_{s}FN(1-{1\over N^{2}})^{1\over2}$, as can be seen from the 
algebra (\ref{su2}). So, in conclusion, the presence of an R--R field has 
condensed the initially free fundamental strings into a static noncommutative 
spherical configuration. Also, we should point out that this solution
corresponds to a string theory derivation of the commutation relations for
fundamental strings in the presence of R--R fields, proposed in \cite{Chu-Ho-Kao}.

This phenomena leads to an interesting question. In the $D$--brane 
situation, the Myers' effect \cite{Myers} tells us that an external 
R--R field can polarize a collection of $D$--branes into a 
noncommutative configuration. The noncommutativity is present at the 
world--volume level, such that there is a background commutative 
spacetime with a noncommutative object made out of many $D$--branes inside.
In the situation described in this section we are dealing with 
$F$--strings. So, even though the situation seems quite analogous to 
the one of the brane system, one also needs to take into account the 
fact that the $F$--strings actually describe gravitons, and we are 
thus led to a situation where the R--R field is building a 
noncommutative object made out of many gravitons. Now, graviton states 
generally correspond to curved spacetimes, however describing small 
fluctuations such as gravitational waves with small amplitudes. If 
one wishes to describe large fluctuations one needs to consider states 
with a semiclassical behavior which would correspond to coherent 
states -- where one would run into the mentioned problems concerning 
exponentiation of nonabelian couplings.

The question still remains on how to interpret a noncommutative 
object made out of many gravitons, inside a background commutative spacetime. 
One speculative possibility is that this could actually correspond to 
a noncommutative background spacetime geometry. Indeed, one could 
imagine that if we put enough gravitons together, the noncommutative 
spherical configuration will grow up to a stage where its curvature 
is actually weak. Then, we would be in a position to expect that this 
large sphere would take up the role of the background spacetime, {\it 
i.e.}, the R--R flux we considered would have created some sort of 
noncommutative background spacetime geometry. It would be quite interesting 
to further study more complex examples of such situations.

\section{Conclusions}

We have seen in this paper how to construct an action for matrix 
string theory in weakly curved background fields. In the process, we 
have also studied its relation to $T$ and $S$ dualities. Such an 
action provides working ground to study multiple interacting strings 
in both NS--NS and R--R backgrounds. For the particular case of an 
R--R background, we have seen how fundamental strings can condense 
into a nonabelian configuration thus building a noncommutative 
stringy configuration. The action we derived also allows for some 
discussion on how background non--linear curvature terms will couple 
to the Matrix theory action in a general curved background. With all 
this in hand, we believe there are quite a few interesting lines for 
future research on this subject. We present some of these lines below.

One possible application of the action we have obtained is to describe 
second quantized superstring theory in general backgrounds with R--R 
fields turned on. This is quite an interesting venue of work, given 
that there has been some recent interest in such ideas, {\it e.g.} 
\cite{Russo-Tseytlin, Metsaev-Tseytlin, Pesando-1, Pesando-2, bvw, Berenstein-Leigh}. Along 
these lines, one should also try to further understand the possible 
background geometry noncommutativity due to the R--R flux, in particular 
it would be quite interesting if some connection to the work in \cite{Jevicki} 
could be done. For this, it could be of some interest to complete further 
examples involving diverse background fields, in order to fully 
explore the nonabelian aspects of curved stringy spacetimes. This 
could be a first step towards the more ambitious goal of constructing 
fully curved noncommutative spacetime geometries from string theory.

The resulting action from our work could also be of use in the study 
of diverse scattering or absorption problems, involving branes, black 
holes, or in the context of the AdS/CFT correspondence 
\cite{Maldacena-conjecture, gkp-2, Witten-AdS1}. It would be quite 
interesting if some work along the lines in \cite{ktr} could be 
accomplished. Probably, from the computation of scattering amplitudes, 
the role of the noncommutative vertex operators would then become more 
translucid (there has been some very recent work on vertex operators in 
\cite{DNP}). This would then be of some interest should it yield 
further insight on how one should exponentiate such operators, in 
order to obtain the full non--linear action. Indeed, one particularly 
interesting use of our work would be to further use this action in order 
to try to infer some new information about Matrix theory in a general 
curved background. This could perhaps be accomplished via a more detailed 
and direct comparison with the Type IIA string theory abelian limit. Some 
work towards such goal has been done in \cite{Douglas-talk, 
Douglas-curved, dko, dos, Douglas-Greene, Douglas-Ooguri, Brax-Wynter, 
Hosomichi-Sugawara, bbg}, and one should try to understand any possible 
relations between those papers and the work presented in here. Probably 
one particularly important relation to understand is the one with the 
work in \cite{Hosomichi-Sugawara}. In order to achieve these goals, some 
research should be done on understanding how to construct the precise 
lift of the matrix string theory action from the cylinder to arbitrary 
Riemann surfaces, and so establish the precise connection to the 
Green--Schwarz action along the lines of \cite{Bonelli-Bonora-Nesti-1, 
Bonelli-Bonora-Nesti-2, Bonelli-Bonora-Nesti-Tomasiello-1}. We hope to 
address some of these questions in the future.

\bigskip
\bigskip

{\bf Acknowledgments:}
I would like to thank Washington Taylor for many discussions and 
critical comments as well as suggestions on the draft of the paper. I would 
also like to thank Lorenzo Cornalba, Jos\'e Mour\~ao and Jo\~ao Nunes for 
helpful discussions and/or comments. The author is supported in part by the 
Funda\c c\~ao para a Ci\^encia e Tecnologia, under the grant Praxis 
XXI BPD-17225/98 (Portugal).

\newpage

\appendix

\section{Supercurrents from Matrix Theory}

We reproduce here the Matrix theory forms of the multipole moments of 
the 11--dimensional supercurrents found in \cite{Dan-Wati-2,Mark-Wati-3}, 
written in 10--dimensional form as in \cite{Mark-Wati-5}. Dropping a 
factor of $1/R$ from each expression, the stress tensor $T^{IJ}$, 
$M2$--brane current $J^{IJK}$ and $M5$--brane current $M^{IJKLMN}$ have 
integrated (monopole) components:

\begin{eqnarray*}
T^{++} &=& \str(\identity) = N\\
T^{+i} &=& -\str(F^{0i})\\
T^{+-} &=& \str(F^{0\mu}F^{0}{}_{\mu} + {1\over 4} F_{\mu\nu} 
F^{\mu\nu} + {i \over 2} \bar{\Theta} \Gamma^{0} D_{0} \Theta)\\
T^{ij} &=& \str(F^{i\mu} F_{\mu}{}^{j} + {i \over 4} \bar{\Theta} 
\Gamma^{i} D_{j} \Theta + {i \over 4} \bar{\Theta} \Gamma^{j} D_{i} \Theta)\\
T^{-i} &=& -\str(F^{0\mu} F_{\mu\nu} F^{\nu i} + {1\over 4} 
F^{0i} F_{\mu\nu} F^{\mu\nu} - {i \over 8} F_{\mu\nu} \bar{\Theta} 
\Gamma^{i} \Gamma^{\mu\nu} D_{0} \Theta \\ 
& & \hspace{0.5in} + {i \over 8} F_{\mu\nu} \bar{\Theta} \Gamma^{0} 
\Gamma^{\mu\nu} D_{i} \Theta - {i \over 4} F_{\mu\nu}\bar{\Theta} 
\Gamma^{\nu} \Gamma^{0i} D^{\mu} \Theta - {1\over 8} \bar{\Theta} 
\Gamma^{0\mu i} \Theta \, \bar{\Theta} \Gamma_{\mu}  \Theta)\\
T^{--} &=& {1\over 4} \str(F_{\mu\nu} F^{\nu\gamma} 
F_{\gamma\delta} F^{\delta\mu} - {1\over 4} F_{\mu\nu} F^{\mu\nu} 
F_{\gamma\delta} F^{\gamma\delta} + i F_{\mu\nu} F_{\gamma\delta} \bar{\Theta} 
\Gamma^{\nu} \Gamma^{\gamma\delta} D^{\mu} \Theta + 
{\cal O}(\Theta^{4}) )\\
J^{+ij} &=& -{1\over 6} \str(F^{ij})\\
J^{+-i} &=& {1\over 6} \str(F^{0\mu} F_{\mu}{}^{i} + {i \over 4} 
\bar{\Theta} \Gamma^{0} D_{i} \Theta - {i \over 4} \bar{\Theta} 
\Gamma^{i} D_{0} \Theta)\\
J^{ijk} &=& {1\over 6} \str( F^{0i} F^{jk} + F^{0j} F^{ki} + F^{0k} 
F^{ij} - {3i \over 4} \bar{\Theta} \Gamma^{0 [ij} D^{k]} \Theta + 
{i \over 4} \bar{\Theta} \Gamma^{ijk} D_{0} \Theta)\\
J^{-ij} &=& {1\over 6} \str(F^{i\mu} F_{\mu\nu} F^{\nu j} + 
{1\over 4} F^{ij} F_{\mu\nu} F^{\mu\nu} - {i \over 8} F_{\mu\nu} 
\bar{\Theta} \Gamma^{j} \Gamma^{\mu\nu} D_{i} \Theta\\ 
& & \hspace{0.5in} + {i \over 8} F_{\mu\nu} \bar{\Theta} 
\Gamma^{i} \Gamma^{\mu\nu} D_{j} \Theta - {i \over 4} F_{\mu\nu} 
\bar{\Theta} \Gamma^{\nu} \Gamma^{ij} D^{\mu} \Theta + {1\over 8} 
\bar{\Theta} \Gamma^{\mu ij} \Theta \bar{\Theta} \Gamma_{\mu} \Theta)\\
M^{+-ijkl} &=& {1\over 12} \str(F^{ij} F^{kl} + F^{ik} F^{lj} + F^{il} 
F^{jk} - i\bar{\Theta} \Gamma^{[ijk}D^{l]} \Theta)\\
M^{-ijklm} &=& -{5\over 4} \str(F^{0[i} F^{jk} F^{lm]} + {i \over 2} 
F^{[0i} \bar{\Theta} \Gamma^{jkl} D^{m]} \Theta).
\end{eqnarray*}

Here, $\str$ denotes a symmetrized trace in which one takes the average 
over all possible orderings of the matrices inside the trace, with 
commutators being treated as a unit block. Time derivatives are taken with 
respect to Minkowski time $t$. Indices $i,j,\ldots$, run from 1 through 9, 
while indices $a,b,\ldots$, run from 0 through 9. In these expressions 
one should use the definitions $F_{0i}=\dot{X}^{i}$ and 
$F_{ij}=i[X^{i},X^{j}]$. A Matrix form for the transverse 5--brane current
components $M^{+ijklm}$ and $M^{ijklmn}$ is not known, and in fact comparison 
with supergravity suggests that these should be zero for any Matrix theory 
configuration.

The higher multipole moments of these currents contain one set of terms which 
are found by including the matrices $X^{k_1},\ldots, X^{k_n}$ into the 
symmetrized trace as well as more complicated spin contributions. We may write 
these as,

\begin{eqnarray}
T^{IJ (i_{1} \cdots i_{k})} &=& \sym (T^{IJ}; X^{i_{1}}, \ldots, X^{i_{k}}) + 
T_{\rm Fermion}^{IJ (i_{1} \cdots i_{k})} \nonumber\\
J^{IJK (i_{1} \cdots i_{k})} &=& \sym (J^{IJK}; X^{i_{1}}, \ldots, X^{i_{k}}) 
+ J_{\rm Fermion}^{IJK (i_{1} \cdots i_{k})} \nonumber\\
M^{IJKLMN (i_{1} \cdots i_{k})} &=& \sym (M^{IJKLMN}; X^{i_{1}}, 
\ldots, X^{i_{k}}) + M_{\rm Fermion}^{IJKLMN (i_{1} \cdots i_{k})},
\nonumber
\end{eqnarray}

\noindent
where some simple examples of the two--fermion contribution to the
first moment terms are,

\begin{eqnarray*}
T_{\rm Fermion}^{+i(j)} &=& -{1\over 8R} \tr(\bar{\Theta} \Gamma^{[0ij]} 
\Theta)\\
T_{\rm Fermion}^{+-(i)} &=& -{1\over 16R} \tr(F_{\mu\nu} \bar{\Theta}  
\gamma^{[\mu\nu i]} \Theta - 4\bar{\Theta} F_{0\mu} \gamma^{[0\mu i]} 
\Theta)\\
T_{\rm Fermion}^{ij(l)} &=& {1\over 8R} \tr(F_{j\mu} \bar{\Theta}  
\gamma^{[\mu il]} \Theta + \bar{\Theta} F_{i\mu} \gamma^{[\mu jl]} 
\Theta)\\
J_{\rm Fermion}^{+ij(k)} &=& {i \over 48R} \tr(\bar{\Theta} 
\Gamma^{[ijk]} \Theta)\\
J_{\rm Fermion}^{+-i(j)} &=& {1\over 48R} \tr(F_{0\mu} \bar{\Theta}  
\gamma^{[\mu ij]} \Theta + \bar{\Theta} F_{i\mu} \gamma^{[\mu 0j]} 
\Theta)\\
M_{\rm Fermion}^{+-ijkl(m)} &=& -{i \over 16R} \str\left(\bar{\Theta} 
F^{[jk}\Gamma^{il]m}\Theta\right).
\end{eqnarray*}  

\noindent
The remaining two--fermion contributions to the first moments and some
four--fermion terms are also determined by the results in
\cite{Mark-Wati-3}. There are also fermionic components of the supercurrent 
which couple to background fermion fields in the supergravity theory. These 
couplings have not been discussed in this paper, but the Matrix theory form 
of the currents is determined in \cite{Mark-Wati-3}.

Finally, there is also a 6--brane current appearing in Matrix theory related 
to nontrivial 11--dimensional background metrics. The components of this 
current as well as its first moments are:

\begin{eqnarray}
S^{+ijklmn} &=& \frac{1}{R} \str\left(F_{[ij} F_{kl} F_{mn]}\right) 
\nonumber\\
S^{+ijklmn(p)} &=& \frac{1}{R} \str\left(F_{[ij} F_{kl} F_{mn]} X_{p} 
- \theta F_{[kl} F_{mn} \gamma_{pqr]} \theta\right) 
\nonumber\\
S^{ijklmnp} &=& \frac{7}{R} \str\left(F_{[ij} F_{kl} F_{mn} \dot{X}_{p]} + 
{\cal O}(\theta^{2},\theta^{4}) \right) 
\nonumber\\
S^{ijklmnp(q)} &=& \frac{7}{R} \str\left(F_{[ij} F_{kl} F_{mn} \dot{X}_{p]} 
X_{q} - \theta \, \dot{X}_{[j} F_{kl} F_{mn} \gamma_{pqr]} \theta + 
{i \over 2} \theta \, F_{[jk} F_{lm}  F_{np} \gamma_{qr]} \theta \right). 
\nonumber
\end{eqnarray}

\eject

\section{DVV Reduction of Matrix Theory Tensors}

The Matrix stress tensor components are:


\bea
T^{++} &=& {1 \over R}\str\left(\identity\right),
\nonumber \\
T^{+i} &=& {1 \over R}\str\left(\dot{X_i}\right),
\nonumber \\
T^{+-} &=& \str\left({1 \over 2R} \dot{X_i} \dot{X_i} - {R M_{P}^{6} \over 
8\pi^{2}} \sum_{i<j} [X^{i},X^{j}]^{2} + {R M_{P}^{6} \over 8\pi^{2}} 
\theta \gamma^i[X^i,\theta]\right),
\nonumber \\
T^{ij} &=& \str\left( {1\over R} \dot{X_i}\dot{X_j} - {R M_{P}^{6} 
\over 4\pi^{2}} [X^{i},X^{k}][X^{k},X^{j}] - {R M_{P}^{6} \over 16\pi^{2}} 
\theta\gamma^i[X_j,\theta] - {R M_{P}^{6} \over 16\pi^{2}} 
\theta\gamma^j[X_i,\theta]\right),
\nonumber \\
T^{-i} &=& \str\left({1\over 2R}\dot{X_i} (\dot{X_j})^{2} - 
{R M_{P}^{6} \over 8\pi^{2}} \dot{X_i} \sum_{j<k} [X^{j},X^{k}]^{2} - 
{R M_{P}^{6} \over 4\pi^{2}} [X^{i},X^{j}][X^{j},X^{k}] \dot{X_k}\right) 
\nonumber\\ 
& 
& - 
\str\left( {R M_{P}^{6} \over 16\pi^{2}} \theta_\alpha 
\dot{X_k}[X_m,\theta_\beta]\right)\{\gamma^k\delta_{im} 
+\gamma^i\delta_{mk} -2\gamma^m\delta_{ki} \}_{\alpha \beta}\nonumber\\ 
& 
& - 
\str\left( {i R^{2} M_{P}^{9} \over 64\pi^{3}} \theta_{\alpha} 
[X^{k},X^{l}] [X_m,\theta_{\beta}]\right)\{ \gamma^{[iklm]} + 
2 \gamma^{[lm]} \delta_{ki} + 4\delta_{ki}\delta_{lm} \}_{\alpha 
\beta}\nonumber\\ 
& 
& + 
\tr\left( {i R^{2} M_{P}^{9} \over 64\pi^{3}} \theta \gamma^{[ki]} \theta \; 
\theta \gamma^k \theta \right),
\nonumber \\
T^{--} &=& \str (
{1\over 4R}(\dot{X^{i}})^{2}(\dot{X^{j}})^{2} + 
{R M_{P}^{6} \over 4\pi^{2}}\dot{X^{i}}\dot{X^{j}}
[X^{i},X^{k}][X^{k},X^{j}] + {R M_{P}^{6} \over 8\pi^{2}}
(\dot{X^{i}})^{2} \sum_{j<k}[X^{j},X^{k}]^{2} 
\nonumber \\ 
& 
& + 
{R^{3} M_{P}^{12} 
\over 64\pi^{4}} [X^{i},X^{j}][X^{j},X^{k}][X^{k},X^{m}][X^{m},X^{i}] - 
{R^{3} M_{P}^{12} \over 64\pi^{4}} \sum_{i<j}[X^{i},X^{j}]^{2} 
\sum_{k<m}[X^{k},X^{m}]^{2} 
\nonumber \\ 
& 
& + 
{\cal O}({\theta^2}) + {\cal O}({\theta^4}) ).
\eea

\noindent
To these components, one should now perform the $T$--duality for the 
$9-11$ flip, followed by the rescalings of world--sheet coordinates, 
background fields and coupling constants. The final result to obtain 
is the explicit form of the previous components of the stress tensor, 
this time in matrix string theory (with $i,j \neq 9$):


\bea \label{mst--}
T^{++} &=& {1 \over 2\pi} \left({\ell_{s} \over R}\right)^{2} 
\int d\sigma d\tau\; \str\left(\identity\right),
\nonumber \\
T^{+i} &=& {1 \over 2\pi} \left({\ell_{s} \over R}\right) 
\int d\sigma d\tau\; \str\left(\dot{X_i}\right),
\nonumber \\
T^{+9} &=& {1 \over 2\pi} \left({\ell_{s} \over R}\right) 
\int d\sigma d\tau\; \str\left(g_{s} \dot{A}\right),
\nonumber \\
T^{+-} &=& {1\over 2\pi} \int d\sigma d\tau\; \str ( {1\over 2}
\dot{X_{i}}^{2} + {1\over 2}(DX^{i})^{2} + {1\over 
2}g_{s}^{2}\dot{A}^{2} - {1\over 2g_{s}^{2}} \sum_{i<j} 
[X^{i},X^{j}]^{2} \nonumber \\
&
& + 
{1\over g_{s}}\theta \gamma_{i}[X^{i},\theta] + 
i\theta \gamma^{9} D\theta ),
\nonumber \\
T^{ij} &=& {1\over 2\pi} \int d\sigma d\tau\; \str ( \dot{X_i}\dot{X_j} 
- D X^{i} D X^{j} - {1\over g_{s}^{2}} [X^{i},X^{k}][X^{k},X^{j}] 
\nonumber \\
&
& - 
{1\over 2 g_{s}} \theta\gamma^i[X_j,\theta] - 
{1\over 2 g_{s}} \theta\gamma^j[X_i,\theta] ),
\nonumber \\
T^{i9} &=& {1\over 2\pi} \int d\sigma d\tau\; \str ( g_{s} 
\dot{X_i}\dot{A} + {i \over g_{s}} [X^{i},X^{k}] D X^{k} - 
{i \over 2} \theta\gamma^{i} D \theta - {1\over 2g_{s}} 
\theta\gamma^{9} [X_i,\theta] ),
\nonumber \\
T^{99} &=& {1\over 2\pi} \int d\sigma d\tau\; \str\left( 
g_{s}^{2} \dot{A}^{2} - (D X^{k})^{2} - i \theta\gamma^{9} D \theta 
\right),
\nonumber \\
T^{-i} &=& {1\over 2\pi} \left({R \over \ell_{s}}\right) 
\int d\sigma d\tau\; \str (
{1\over 2}\dot{X_i} (\dot{X_j})^{2} + {1\over 2} g_{s}^{2} 
\dot{X_{i}}\dot{A}^{2} - {1\over 2 g_{s}^{2}} \dot{X_i} \sum_{j<k} 
[X^{j},X^{k}]^{2} + {1\over 2} \dot{X_{i}}(D X^{j})^{2} 
\nonumber \\
&
& - 
{1\over g_{2}^{2}} [X^{i},X^{j}][X^{j},X^{k}] \dot{X_k} - 
D X^{i} D X^{k} \dot{X_k} + i [X^{i},X^{j}] D X^{j} \dot{A} 
\nonumber \\ 
& 
& - 
{1\over 2g_{s}} \theta_\alpha 
\dot{X_k}[X_j,\theta_\beta] \{\gamma^k\delta_{ij} 
+\gamma^i\delta_{jk} -2\gamma^j\delta_{ki} \}_{\alpha \beta} - 
{1\over 2} \dot{A} \theta \gamma^{9} [X_i,\theta] + 
i \dot{X_i} \theta \gamma^{9} D \theta 
\nonumber\\ 
& 
& - 
{i \over 2} g_{s} \dot{A} \theta \gamma^{i} D \theta - 
{i \over 4g_{s}^{2}} \theta_{\alpha} [X^{k},X^{j}] [X^{l},\theta_{\beta}] 
\{ \gamma^{[ikjl]} + 2 \gamma^{[jl]} \delta_{ki} + 
4\delta_{ki}\delta_{jl} \}_{\alpha \beta} 
\nonumber\\ 
&  
& - 
{1 \over 2g_{s}} \theta_{\alpha} D X^{k} [X^{j},\theta_{\beta}] 
\{ \gamma^{[ik9j]} + \gamma^{[9j]} \delta_{ki} \}_{\alpha \beta} + 
{1 \over 4g_{s}} \theta_{\alpha} [X^{k},X^{j}] D \theta_{\beta} 
\{ \gamma^{[ikj9]} + 2 \gamma^{[j9]} \delta_{ki} \}_{\alpha \beta} 
\nonumber \\ 
& 
& - 
i D X^{i} \theta D \theta + 
{i \over 2} \left({R\over \ell_{s}}\right) ( \theta \gamma^{[ki]} \theta \; 
\theta \gamma^k \theta + \theta \gamma^{[9i]} \theta \; 
\theta \gamma^9 \theta) ),
\nonumber \\
T^{-9} &=& {1\over 2\pi} \left({R \over \ell_{s}}\right) 
\int d\sigma d\tau\; \str (
{1\over 2} g_{s}\dot{A} (\dot{X^{i}})^{2} + {1\over 2} g_{s}^{3}\dot{A}^{3} 
- {1\over 2 g_{s}} \dot{A} \sum_{i<j} [X^{i},X^{j}]^{2} 
\nonumber \\
&
& - 
{i \over g_{s}} D X^{i} [X^{i},X^{j}] \dot{X^{j}} - 
{1\over 2} g_{s} (D X^{i})^{2} \dot{A} - 
{1 \over 2 g_{s}} \dot{X^{i}} \theta 
\gamma^{9} [X^{i},\theta] + \dot{A} \theta \gamma^{i} [X^{i},\theta] 
\nonumber \\ 
& 
& - 
{i \over 2} \dot{X^{i}} \theta \gamma^{i} D\theta - {i \over 4 
g_{s}^{2}} [X^{i},X^{j}] \theta \gamma^{[9ijk]} [X^{k},\theta] + 
{1 \over 2 g_{s}} \theta_{\alpha} D X^{i} [X^{j},\theta_{\beta}] 
\{ \gamma^{[ij]} + 2\delta_{ij} \}_{\alpha \beta} 
\nonumber \\ 
& 
& + 
{i \over 2} D X^{i} \theta \gamma^{[i9]} D \theta + 
{i \over 2} \left({R\over \ell_{s}}\right) \theta \gamma^{[i9]} \theta \; 
\theta \gamma^{i} \theta ),
\nonumber \\
T^{--} &=& {1\over 2\pi} \left({R \over \ell_{s}}\right)^{2} 
\int d\sigma d\tau\; \str (
{1\over 4}(\dot{X^{i}})^{2}(\dot{X^{j}})^{2} + 
{1\over 2} g_{s}^{2} \dot{A}^{2}(\dot{X^{i}})^{2} + 
{1\over 4} g_{s}^{4} \dot{A}^{4} 
\nonumber \\ 
& 
& + 
{1\over g_{s}^{2}}\dot{X^{i}}\dot{X^{j}}
[X^{i},X^{k}][X^{k},X^{j}] - 
2 i \dot{X^{i}}\dot{A}
[X^{i},X^{k}] D X^{k} + 
{1\over 2} g_{s}^{2} \dot{A}^{2} (D X^{i})^{2} 
\nonumber \\ 
& 
& + 
\dot{X^{i}}\dot{X^{j}} D X^{i} D X^{j} + 
{1\over 2 g_{s}^{2}} (\dot{X^{i}})^{2} \sum_{j<k}[X^{j},X^{k}]^{2} - 
{1\over 2} (\dot{X^{i}})^{2} (D X^{j})^{2} + 
{1\over 2} \dot{A}^{2} \sum_{i<j}[X^{i},X^{j}]^{2} 
\nonumber \\ 
& 
& + 
{1\over 4 g_{s}^{4}} [X^{i},X^{j}][X^{j},X^{k}][X^{k},X^{m}][X^{m},X^{i}] + 
{1\over g_{s}^{2}} D X^{i} D X^{j} [X^{i},X^{k}][X^{k},X^{j}] 
\nonumber \\ 
& 
& - 
{1\over 4 g_{s}^{4}} \sum_{i<j}[X^{i},X^{j}]^{2} \sum_{k<m}[X^{k},X^{m}]^{2} + 
{1\over 2 g_{s}^{2}} (D X^{i})^{2} \sum_{j<k}[X^{j},X^{k}]^{2} + 
{1\over 4} (D X^{i})^{2} (D X^{j})^{2} 
\nonumber \\ 
& 
& + 
{\cal O}({\theta^2}) + {\cal O}({\theta^4}) ).
\eea

\noindent
Finally, the free string limit can be taken. The result for the 
conformal field theory limit of the matrix string stress tensor is:


\bea
\lim_{g_{s}\to 0} T^{++} &=& {1 \over 2\pi} \left({\ell_{s} \over 
R}\right)^{2} \int d\sigma d\tau\; \str\left(\identity\right),
\nonumber \\
\lim_{g_{s}\to 0} T^{+i} &=& {1 \over 2\pi} \left({\ell_{s} \over 
R}\right) \int d\sigma d\tau\; \str\left(\dot{X_i}\right),
\nonumber \\
\lim_{g_{s}\to 0} T^{+9} &=& 0,
\nonumber \\
\lim_{g_{s}\to 0} T^{+-} &=& {1\over 2\pi} \int d\sigma d\tau\; 
\str\left( {1\over 2}\dot{X_{i}}^{2} + {1\over 2}(\partial X^{i})^{2} + 
i\theta \gamma^{9}\partial \theta \right),
\nonumber \\
\lim_{g_{s}\to 0} T^{ij} &=& {1\over 2\pi} \int d\sigma d\tau\; \str 
\left( \dot{X_i}\dot{X_j} - \partial X^{i} \partial X^{j}\right),
\nonumber \\
\lim_{g_{s}\to 0} T^{i9} &=& {1\over 2\pi} \int d\sigma d\tau\; \str 
\left(- {i \over 2} \theta\gamma^{i} \partial \theta \right),
\nonumber \\
\lim_{g_{s}\to 0} T^{99} &=& {1\over 2\pi} \int d\sigma d\tau\; \str 
\left(- (\partial X^{i})^{2} - i \theta\gamma^{9} \partial \theta 
\right),
\nonumber \\
\lim_{g_{s}\to 0} T^{-i} &=& {1\over 2\pi} \left({R \over \ell_{s}}\right) 
\int d\sigma d\tau\; \str (
{1\over 2}\dot{X_i} (\dot{X_j})^{2} + {1\over 2} \dot{X_{i}}(\partial 
X^{j})^{2} - \partial X^{i} \partial X^{k} \dot{X_k} 
\nonumber \\ 
& 
& +
i \dot{X_i} \theta \gamma^{9} \partial \theta - 
i \partial X^{i} \theta \partial \theta + {\cal O}\left({R\over \ell_{s}} \right)),
\nonumber \\
\lim_{g_{s}\to 0} T^{-9} &=& {1\over 2\pi} \left({R \over \ell_{s}}\right) 
\int d\sigma d\tau\; \str \left(
- {i \over 2} \dot{X^{i}} \theta \gamma^{i} \partial\theta 
+ {i \over 2} \partial X^{i} \theta \gamma^{[i9]} \partial\theta + 
{\cal O}\left({R\over \ell_{s}} \right) \right),
\nonumber \\
\lim_{g_{s}\to 0} T^{--} &=& {1\over 2\pi} \left({R \over \ell_{s}}\right)^{2} 
\int d\sigma d\tau\; \str (
{1\over 4}(\dot{X^{i}})^{2}(\dot{X^{j}})^{2} + 
\dot{X^{i}}\dot{X^{j}} \partial X^{i} \partial X^{j} - 
{1\over 2} (\dot{X^{i}})^{2} (\partial X^{j})^{2} 
\nonumber \\ 
& 
&  +
{1\over 4} (\partial X^{i})^{2} (\partial X^{j})^{2} + 
{\cal O}({\theta^2}) + {\cal O}({\theta^4}) ).
\eea

The next terms we look at are the zeroth moments of the components 
of the Matrix membrane current. These components are:


\bea
J^{+ij} &=& - {i M_{P}^{3} \over 12\pi} \str\left([X^{i},X^{j}]\right),
\nonumber \\
J^{+-i} &=& \str\left( {i M_{P}^{3} \over 12\pi}[X^{i},X^{j}] \dot{X^{j}} - 
{R M_{P}^{6} \over 48\pi^{2}} \theta[X^{i},\theta] + 
{R M_{P}^{6} \over 96\pi^{2}} \theta \gamma^{[ki]} [X^{k}, 
\theta]\right),
\nonumber \\
J^{ijk} &=& - \str\left( {i M_{P}^{3} \over 12\pi} \dot{X^{i}}[X^{j},X^{k}] + 
{i M_{P}^{3} \over 12\pi} \dot{X^{j}}[X^{k},X^{i}] + 
{i M_{P}^{3} \over 12\pi} \dot{X^{k}}[X^{i},X^{j}] - {R M_{P}^{6} \over 
96\pi^{2}} \theta \gamma^{[ijkl]}[X_l,\theta]\right),
\nonumber \\
J^{-ij} &=& \str ( 
{i M_{P} ^{3} \over 12\pi} \dot{X^{i}} \dot{X^{k}} [X^{k},X^{j}] - 
{i M_{P} ^{3} \over 12\pi} \dot{X^{j}} \dot{X^{k}} [X^{k},X^{i}] - 
{i M_{P} ^{3} \over 24\pi} (\dot{X^{k}})^{2} [X^{i},X^{j}] 
\nonumber \\ 
& 
& - 
{i R^{2} M_{P} ^{9} \over 96\pi^{3}} [X^{i},X^{j}] \sum_{k<l} 
[X^{k},X^{l}]^{2} - 
{i R^{2} M_{P} ^{9} \over 48\pi^{3}} [X^{i},X^{k}] [X^{k},X^{l}] 
[X^{l},X^{j}] ) 
\nonumber\\
& 
& + 
{R M_{P}^{6} \over 96\pi^{2}} \str\left(\theta_{\alpha} 
\dot{X^{k}}[X^{m},\theta_{\beta}]\right) \{ \gamma^{[kijm]} + 
\gamma^{[jm]} \delta_{ki} - \gamma^{[im]} \delta_{kj} + 2 \delta_{jm} 
\delta_{ki} - 2 \delta_{im} \delta_{kj}\}_{\alpha \beta}
\nonumber\\
& 
& + 
{i R^{2} M_{P}^{9} \over 64\pi^{3}} \str\left(\theta_{\alpha} 
[X^{k},X^{l}][X^{m},\theta_{\beta}]\right)\{\gamma^{[jkl]} 
\delta_{mi} - \gamma^{[ikl]} \delta_{mj} + 2 \gamma^{[lij]} \delta_{km} 
+ 2 \gamma^{l} \delta_{jk} \delta_{im} 
\nonumber\\
& 
& \hspace{1in} - 
2 \gamma^{l} \delta_{ik} \delta_{jm} + 2 \gamma^{j} \delta_{il} 
\delta_{km} - 2 \gamma^{i} \delta_{jl} \delta_{km}\}_{\alpha \beta}
\nonumber\\ 
& 
& + 
{i R^{2} M_{P}^{9} \over 384\pi^{3}} \str\left(\theta 
\gamma^{[kij]} \theta \; \theta \gamma^k \theta - 
\theta \gamma^{[ij]} \theta \; \theta \theta\right).
\eea

\noindent
To these components we now perform the $T$--duality for the 
$9-11$ flip, followed by the rescalings of world--sheet coordinates, 
background fields and coupling constants. One obtains the explicit form of 
the previous components of the membrane current, in matrix string theory 
(with $i,j,k \neq 9$):


\bea
J^{+ij} &=& {1\over 2\pi} \left({\ell_{s}\over R}\right) 
\int d\sigma d\tau\; \str\left(-{i \over 6 g_{s}} [X^{i},X^{j}]\right),
\nonumber \\
J^{+i9} &=& {1\over 2\pi} \left({\ell_{s}\over R}\right) 
\int d\sigma d\tau\; \str\left(
-{1\over 6} D X^{i}\right),
\nonumber \\
J^{+-i} &=& {1\over 2\pi} \int d\sigma d\tau\; \str (
{i \over 6 g_{s}}[X^{i},X^{j}] \dot{X^{j}} + 
{1\over 6} g_{s} \dot{A} D X^{i} - 
{1\over 6 g_{s}} \theta[X^{i},\theta] 
\nonumber \\ 
& 
& + 
{1\over 12 g_{s}} \theta \gamma^{[ki]} [X^{k}, \theta] + 
{i \over 12} \theta \gamma^{[9i]} D \theta ),
\nonumber \\
J^{+-9} &=& {1\over 2\pi} \int d\sigma d\tau\; \str\left(
- {1\over 6} \dot{X^{i}} D X^{i} - {i \over 6} \theta D \theta + 
{1 \over 12 g_{s}} \theta \gamma^{[i9]} [X^{i},\theta]\right),
\nonumber \\
J^{ijk} &=& {1\over 2\pi} \int d\sigma d\tau\; \str (
- {i \over 6 g_{s}} \dot{X^{i}}[X^{j},X^{k}] - 
{i \over 6 g_{s}} \dot{X^{j}}[X^{k},X^{i}] - 
{i \over 6 g_{s}} \dot{X^{k}}[X^{i},X^{j}] 
\nonumber \\ 
& 
& + 
{1\over 12 g_{s}} \theta \gamma^{[ijkl]}[X_l,\theta] + 
{i \over 12} \theta \gamma^{[ijk9]} D \theta ),
\nonumber \\
J^{ij9} &=& {1\over 2\pi} \int d\sigma d\tau\; \str\left(
- {1\over 6} \dot{X^{i}}D X^{j} + {1\over 6} \dot{X^{j}}D X^{i} - 
{i \over 6} \dot{A}[X^{i},X^{j}] + 
{1\over 12 g_{s}} \theta \gamma^{[ij9l]}[X_l,\theta] \right),
\nonumber \\
J^{-ij} &=& {1\over 2\pi} \left({R \over \ell_{s}} \right) 
\int d\sigma d\tau\; \str (
{i \over 6 g_{s}} \dot{X^{i}} \dot{X^{k}} [X^{k},X^{j}] - 
{i \over 6 g_{s}} \dot{X^{j}} \dot{X^{k}} [X^{k},X^{i}] - 
{g_{s} \over 6} \dot{A} \dot{X^{i}} D X^{j} 
\nonumber \\ 
& 
& + 
{g_{s} \over 6} \dot{A} \dot{X^{j}} D X^{i} - 
{i \over 12 g_{s}} (\dot{X^{k}})^{2} [X^{i},X^{j}] - 
{i \over 12} g_{s} \dot{A}^{2} [X^{i},X^{j}] - 
{i \over 12 g_{s}^{3}} [X^{i},X^{j}] \sum_{k<l} [X^{k},X^{l}]^{2} 
\nonumber \\ 
& 
& + 
{i \over 12 g_{s}} [X^{i},X^{j}] (D X^{k})^{2} - 
{i \over 6 g_{s}^{3}} [X^{i},X^{k}] [X^{k},X^{l}] [X^{l},X^{j}] - 
{i \over 6 g_{s}} D X^{i} D X^{k} [X^{k},X^{j}] 
\nonumber \\ 
& 
& + 
{i \over 6 g_{s}} D X^{j} D X^{k} [X^{k},X^{i}] + 
{1\over 12 g_{s}} \theta_{\alpha} 
\dot{X^{k}}[X^{m},\theta_{\beta}] \{ \gamma^{[kijm]} + 
\gamma^{[jm]} \delta_{ki} - \gamma^{[im]} \delta_{kj} 
\nonumber \\ 
& 
& \hspace{1in} + 
2 \delta_{jm} \delta_{ki} - 2 \delta_{im} \delta_{kj}\}_{\alpha \beta}
\nonumber \\ 
& 
& + 
{1\over 12} \dot{A} \theta \gamma^{[9ijm]} [X^{m},\theta] + 
{i \over 12} \theta_{\alpha} \dot{X^{k}} D \theta_{\beta} \{ 
\gamma^{[kij9]} + \gamma^{[j9]} \delta_{ki} - \gamma^{[i9]} 
\delta_{kj} \}_{\alpha \beta}
\nonumber\\
& 
& + 
{i \over 4 g_{s}^{2}} \theta_{\alpha} 
[X^{k},X^{l}][X^{m},\theta_{\beta}] \{\gamma^{[jkl]} 
\delta_{mi} - \gamma^{[ikl]} \delta_{mj} + 2 \gamma^{[lij]} \delta_{km} 
+ 2 \gamma^{l} \delta_{jk} \delta_{im} 
\nonumber\\
& 
& \hspace{1in} - 
2 \gamma^{l} \delta_{ik} \delta_{jm} + 2 \gamma^{j} \delta_{il} 
\delta_{km} - 2 \gamma^{i} \delta_{jl} \delta_{km}\}_{\alpha \beta}
\nonumber\\ 
& 
& +
{1\over 2 g_{s}} \theta_{\alpha} D X^{k} [X^{m},\theta_{\beta}] 
\{ \gamma^{[jk9]} \delta_{mi} - \gamma^{[ik9]} \delta_{mj} + 
\gamma^{[9ij]} \delta_{km} + 
\gamma^{9} \delta_{jk} \delta_{im} - 
\gamma^{9} \delta_{ik} \delta_{jm} \}_{\alpha \beta} 
\nonumber \\ 
& 
& - 
{i \over 2} \theta_{\alpha} D X^{l} D \theta_{\beta} \{\gamma^{[lij]} 
+ \gamma^{j} \delta_{il} - \gamma^{i} \delta_{jl} \}_{\alpha \beta} 
\nonumber \\
& 
& + 
{i \over 12} \left({R\over \ell_{s}} \right) \left(\theta 
\gamma^{[kij]} \theta \; \theta \gamma^{k} \theta + \theta 
\gamma^{[9ij]} \theta \; \theta \gamma^{9} \theta - 
\theta \gamma^{[ij]} \theta \; \theta \theta\right) ),
\nonumber \\
J^{-i9} &=& {1\over 2\pi} \left({R \over \ell_{s}} \right) 
\int d\sigma d\tau\; \str (
{1\over 6} \dot{X^{i}} \dot{X^{k}} D X^{k} - 
{i \over 6} \dot{A} \dot{X^{k}} [X^{k},X^{i}] + 
{1\over 12} g_{s}^{2} \dot{A}^{2} D X^{i} 
\nonumber \\ 
& 
& - 
{1\over 12} (\dot{X^{k}})^{2} D X^{i} - 
{1\over 12 g_{s}^{2}} D X^{i} \sum_{k<l} [X^{k},X^{l}]^{2} - 
{1\over 12} D X^{i} (D X^{k})^{2} 
\nonumber \\ 
& 
& - 
{1\over 6 g_{s}^{2}} [X^{i},X^{k}] [X^{k},X^{l}] D X^{l} + 
{1\over 12 g_{s}} \theta_{\alpha} 
\dot{X^{k}}[X^{m},\theta_{\beta}] \{ \gamma^{[ki9m]} + 
\gamma^{[9m]} \delta_{ki} \}_{\alpha \beta}
\nonumber \\ 
& 
& - 
{1\over 12} \dot{A} \theta_{\alpha} 
[X^{m},\theta_{\beta}] \{ \gamma^{[im]} + 2 \delta_{im} \}_{\alpha 
\beta} + 
{i \over 6} \dot{X^{i}} \theta D \theta - 
{i \over 12} \dot{A} \theta \gamma^{[i9]} D \theta 
\nonumber \\ 
& 
& + 
{i \over 4 g_{s}^{2}} \theta_{\alpha} 
[X^{k},X^{l}][X^{m},\theta_{\beta}] \{\gamma^{[9kl]} \delta_{mi} + 
2 \gamma^{[li9]} \delta_{km} + 2 \gamma^{9} \delta_{il} \delta_{km} 
\}_{\alpha \beta}
\nonumber\\
& 
& - 
{1\over 2 g_{s}} D X^{k} \theta \gamma^{k} [X^{i},\theta] - 
{1\over 2 g_{s}} D X^{k} \theta \gamma^{i} [X^{k},\theta] 
\nonumber \\ 
& 
& + 
{i \over 4 g_{s}} \theta_{\alpha} 
[X^{k},X^{l}] D \theta_{\beta} \{ \gamma^{[ikl]} + 2 
\gamma^{l} \delta_{ik} \}_{\alpha \beta} - 
i D X^{i}\theta \gamma^{9} D \theta
\nonumber \\
& 
& + 
{i \over 12} \left({R\over \ell_{s}} \right)\left(\theta 
\gamma^{[ki9]} \theta \; \theta \gamma^{k} \theta - 
\theta \gamma^{[i9]} \theta \; \theta \theta\right) ).
\eea

\noindent
The free string limit can now be taken. The result for the 
conformal field theory limit of the matrix string membrane current is:


\bea
\lim_{g_{s}\to 0} J^{+ij} &=& 0,
\nonumber \\
\lim_{g_{s}\to 0} J^{+i9} &=& {1\over 2\pi} \left({\ell_{s}\over R}\right) 
\int d\sigma d\tau\; \str\left(-{1\over 6} \partial X^{i}\right),
\nonumber \\
\lim_{g_{s}\to 0} J^{+-i} &=& {1\over 2\pi} \int d\sigma d\tau\; 
\str\left({i \over 12} \theta \gamma^{[9i]} \partial \theta \right),
\nonumber \\
\lim_{g_{s}\to 0} J^{+-9} &=& {1\over 2\pi} \int d\sigma d\tau\; \str\left(
- {1\over 6} \dot{X^{i}} \partial X^{i} - {i \over 6} \theta 
\partial \theta \right),
\nonumber \\
\lim_{g_{s}\to 0} J^{ijk} &=& {1\over 2\pi} 
\int d\sigma d\tau\; \str\left( 
{i \over 12} \theta \gamma^{[ijk9]} \partial \theta \right),
\nonumber \\
\lim_{g_{s}\to 0} J^{ij9} &=& {1\over 2\pi} \int d\sigma d\tau\; 
\str\left( - {1\over 6} \dot{X^{i}}\partial X^{j} + 
{1\over 6} \dot{X^{j}}\partial X^{i} \right),
\nonumber \\
\lim_{g_{s}\to 0} J^{-ij} &=& {1\over 2\pi} \left({R \over \ell_{s}} \right) 
\int d\sigma d\tau\; \str (
{i \over 12} \theta_{\alpha} \dot{X^{k}} \partial \theta_{\beta} \{ 
\gamma^{[kij9]} + \gamma^{[j9]} \delta_{ki} - \gamma^{[i9]} 
\delta_{kj} \}_{\alpha \beta}
\nonumber\\
& 
& - 
{i \over 2} \theta_{\alpha} \partial X^{l} \partial \theta_{\beta} 
\{\gamma^{[lij]} + \gamma^{j} \delta_{il} - \gamma^{i} \delta_{jl} 
\}_{\alpha \beta} + {\cal O}\left({R\over \ell_{s}} \right) ),
\nonumber \\
\lim_{g_{s}\to 0} J^{-i9} &=& {1\over 2\pi} \left({R \over \ell_{s}} \right) 
\int d\sigma d\tau\; \str (
{1\over 6} \dot{X^{i}} \dot{X^{k}} \partial X^{k} - 
{1\over 12} (\dot{X^{k}})^{2} \partial X^{i} - 
{1\over 12} \partial X^{i} (\partial X^{k})^{2} 
\nonumber \\ 
& 
& + 
{i \over 6} \dot{X^{i}} \theta \partial \theta - 
i \partial X^{i}\theta \gamma^{9} \partial \theta + 
{\cal O}\left({R\over \ell_{s}} \right)).
\eea

Next, we look at the zeroth moments of the components of the Matrix 
5--brane current. Explicitly, these components are:


\bea
M^{+-ijkl} &=& \str(
- {R M_{P}^{6} \over 48\pi^{2}} [X^{i},X^{j}][X^{k},X^{l}]
- {R M_{P}^{6} \over 48\pi^{2}} [X^{i},X^{k}][X^{l},X^{j}] 
\nonumber \\ 
& 
& - 
{R M_{P}^{6} \over 48\pi^{2}} [X^{i},X^{l}][X^{j},X^{k}]
+{R M_{P}^{6} \over 48\pi^{2}} \theta \gamma^{[jkl}[X^{i]},\theta] ),
\nonumber \\
M^{-ijklm} &=& \str ( 
-{5R M_{P}^{6} \over 16\pi^{2}}\dot{X}^{[i}[X^{j},X^{k}][X^{l},X^{m]}]
-{5R M_{P}^{6} \over 48\pi^{2}}\theta\dot{X}^{[i}\gamma^{jkl}[X^{m]},\theta] 
\nonumber \\ 
& 
& - 
{5i R^{2} M_{P}^{9} \over 192\pi^{3}}
\theta [X^{[i},X^{j}]\gamma^{klm]}\gamma^{n} [X^{n},\theta] ).
\eea

\noindent
To these components one now performs the $T$--duality for the 
$9-11$ flip, followed by the rescalings of world--sheet coordinates, 
background fields and coupling constants. We then obtain the explicit form of 
the previous components of the matrix string theory 5--brane current 
(with $i,j,k,l,m \neq 9$):


\bea
M^{+-ijkl} &=& {1\over 2\pi} \int d\sigma d\tau\; \str (
- {1\over 12 g_{s}^{2}} [X^{i},X^{j}][X^{k},X^{l}]
- {1\over 12 g_{s}^{2}} [X^{i},X^{k}][X^{l},X^{j}] 
\nonumber \\ 
& 
& - 
{1\over 12 g_{s}^{2}} [X^{i},X^{l}][X^{j},X^{k}]
+{1\over 6 g_{s}} \theta \gamma^{[jkl}[X^{i]},\theta] ),
\nonumber \\
M^{+-ijk9} &=& {1\over 2\pi} \int d\sigma d\tau\; \str (
{i \over 2 g_{s}} D X^{k}[X^{i},X^{j}]
-{i \over 2 g_{s}} D X^{j}[X^{i},X^{k}] + 
{i \over 2 g_{s}} D X^{i}[X^{j},X^{k}] 
\nonumber \\ 
& 
& + 
{i \over 6} \theta \gamma^{[ikj]}D \theta - 
{1\over 6 g_{s}} 
\left( \theta \gamma^{9[kj}[X^{i]},\theta] + \theta 
\gamma^{[i|9|j}[X^{k]},\theta] + \theta \gamma^{[ik|9|}[X^{j]},\theta] 
\right) ),
\nonumber \\
M^{-ijklm} &=& {1\over 2\pi} \left({R \over \ell_{s}}\right) 
\int d\sigma d\tau\; \str (
-{5\over 4 g_{s}^{2}}\dot{X}^{[i}[X^{j},X^{k}][X^{l},X^{m]}]
-{5\over 6 g_{s}}\theta\dot{X}^{[i}\gamma^{jkl}[X^{m]},\theta] 
\nonumber \\ 
& 
& - 
{5i \over 12 g_{s}^{2}}
\theta [X^{[i},X^{j}]\gamma^{klm]}\gamma^{n} [X^{n},\theta] + 
{5\over 12 g_{s}}
\theta [X^{[i},X^{j}]\gamma^{klm]} \gamma^{9} D \theta ),
\nonumber \\
M^{-ijkl9} &=& {1\over 2\pi} \left({R \over \ell_{s}}\right) 
\int d\sigma d\tau\; \str (
{5 i \over g_{s}}\dot{X}^{[i}[X^{j},X^{k}] D X^{l]} + 
{5\over 4 g_{s}}\dot{A} [X^{[i},X^{j}][X^{l},X^{k]}]
\nonumber \\ 
& 
& - 
{5 i \over 6}\theta\dot{X}^{[i}\gamma^{jkl]} D \theta + 
{5\over 6} \dot{A} \theta \gamma^{[ijl} [X^{k]},\theta] + 
{5\over 2 g_{s}} \theta\dot{X}^{[i}\gamma^{|9|kl} [X^{j]},\theta] 
\nonumber \\ 
& 
& - 
{5i \over 4 g_{s}^{2}} \theta [X^{[i},X^{j}]\gamma^{kl]9} \gamma^{n} 
[X^{n},\theta] + {5\over 4 g_{s}} \theta [X^{[i},X^{j}]\gamma^{kl]9} 
\gamma^{9} D \theta 
\nonumber \\ 
& 
& - 
{5\over 6 g_{s}} \theta D X^{[i}\gamma^{jlk]} \gamma^{n} 
[X^{n},\theta] - {5i \over 6} \theta D X^{[i}\gamma^{jlk]} 
\gamma^{9} D \theta ).
\eea

\noindent
One can now take the free string limit. The result for the conformal 
field theory limit of the matrix string 5--brane current is:


\bea
\lim_{g_{s}\to 0} M^{+-ijkl} &=& 0,
\nonumber \\
\lim_{g_{s}\to 0} M^{+-ijk9} &=& {1\over 2\pi} \int d\sigma d\tau\; 
\str\left({i \over 6} \theta \gamma^{[ikj]}\partial \theta \right),
\nonumber \\
\lim_{g_{s}\to 0} M^{-ijklm} &=& 0,
\nonumber \\
\lim_{g_{s}\to 0} M^{-ijkl9} &=& {1\over 2\pi} 
\left({R \over \ell_{s}}\right) \int d\sigma d\tau\; \str\left(
- {5 i \over 6}\theta\dot{X}^{[i}\gamma^{jkl]} \partial \theta 
- {5i \over 6} \theta \partial X^{[i}\gamma^{jlk]} 
\gamma^{9} \partial \theta \right).
\eea

Finally, we look at are the zeroth moments of the components of the Matrix 
6--brane current (related to nontrivial $11$--dimensional background 
metrics), which are given explicitly by:


\bea
S^{+ijklmn} &=& - {i R^{2} M_{P}^{9} \over 8\pi^{3}} \str\left( 
[X^{[i},X^{j}][X^{k},X^{l}][X^{m},X^{n]}] \right),
\nonumber \\
S^{ijklmnp} &=& \str\left( -{7i R^{2} M_{P}^{9} \over 8\pi^{3}} 
[X^{[i},X^{j}][X^{k},X^{l}][X^{m},X^{n}]\dot{X}^{p]} + 
{\cal O}(\theta^{2},\theta^{4}) \right).
\eea

\noindent
To these components we now perform $T$--duality for the 
$9-11$ flip, followed by the rescalings of world--sheet coordinates, 
background fields and coupling constants. We obtain the explicit form of 
the previous components of the matrix string theory 6--brane current 
(with $i,j,k,l,m,n \neq 9$):


\bea
S^{+ijklmn} &=& {1\over 2\pi} \left({R \over \ell_{s}}\right) 
\int d\sigma d\tau\; \str\left(- {i \over g_{s}^{3}} 
[X^{[i},X^{j}][X^{k},X^{l}][X^{m},X^{n]}] \right),
\nonumber \\
S^{+ijklm9} &=& {1\over 2\pi} \left({R \over \ell_{s}}\right) 
\int d\sigma d\tau\; \str\left(
- {6\over g_{s}^{2}} [X^{[i},X^{j}][X^{k},X^{l}] D X^{m]} \right),
\nonumber \\
S^{ijklmnp} &=& {1\over 2\pi} \left({R \over \ell_{s}}\right)^{2} 
\int d\sigma d\tau\; \str\left(
-{7i \over g_{s}^{3}} 
[X^{[i},X^{j}][X^{k},X^{l}][X^{m},X^{n}]\dot{X}^{p]} + 
{\cal O}(\theta^{2},\theta^{4}) \right),
\nonumber \\
S^{ijklmn9} &=& {1\over 2\pi} \left({R \over \ell_{s}}\right)^{2} 
\int d\sigma d\tau\; \str ( -{7i \over g_{s}^{2}} 
[X^{[i},X^{j}][X^{k},X^{l}][X^{m},X^{n]}]\dot{A} 
\nonumber \\ 
& 
& - 
{42\over g_{s}^{2}} D X^{[j}][X^{k},X^{l}][X^{m},X^{n]}]\dot{X}^{i]} + 
{\cal O}(\theta^{2},\theta^{4}) ).
\eea

\noindent
One can now take the free string limit. The result for the conformal 
field theory limit of the matrix string 6--brane current is:


\bea
\lim_{g_{s}\to 0} S^{+ijklmn} &=& 0,
\nonumber \\
\lim_{g_{s}\to 0} S^{+ijklm9} &=& 0,
\nonumber \\
\lim_{g_{s}\to 0} S^{ijklmnp} &=& 0,
\nonumber \\
\lim_{g_{s}\to 0} S^{ijklmn9} &=& 0.
\eea

\newpage

\bibliographystyle{plain}

\end{document}